%% file: paper.tex
\documentclass[sigplan,10pt]{acmart}  %

\usepackage{color,graphicx,amsmath,amsthm,array, subcaption}
\usepackage{boxedminipage,multirow,endnotes,balance}
\usepackage{pdfpages}
\usepackage{rotating}
\usepackage{colortbl}
\usepackage{transparent}
\usepackage{appendix}
\usepackage{multirow}
\usepackage{relsize}
\usepackage{booktabs}
\usepackage{multirow}
\usepackage[font={small,bf}, skip=1pt]{caption}
\usepackage{cancel}
\usepackage{xcolor,times}
\usepackage[linesnumbered,ruled,vlined]{algorithm2e}
\usepackage{soul}

\usepackage{tikz}
\DeclareRobustCommand\numcircledtikz[1]{\tikz[baseline=(char.base)]{
    \node[shape=circle,draw,fill,inner sep=1pt] (char)
    {\textcolor{white}{#1}};}}

  {\begin{list}{$\bullet$}%
     {\setlength{\parsep}{0pt}%
      \setlength{\topsep}{0pt}%
      \setlength{\itemsep}{0pt}}}%
  {\end{list}}
  
\makeatletter
\let\@authorsaddresses\@empty %
\newcommand\blfootnote[1]{%
    \begingroup
    \renewcommand\thefootnote{}\footnote{#1}%
    \addtocounter{footnote}{-1}%
    \endgroup
}
\global\def\section{\@startsection {section}{1}{\z@}%
                                   {-1.5ex \@plus -0.8ex \@minus -.1ex}%
                                   {0.6ex \@plus.2ex}%
                                   {\normalfont\bfseries\scshape\fontsize{11}{13}\selectfont}}
\global\def\subsection{\@startsection{subsection}{2}{\z@}%
                                     {-1.25ex\@plus -0.8ex \@minus -.1ex}%
                                     {0.3ex \@plus .1ex}%
                                     {\normalfont\bfseries\fontsize{10}{12}\selectfont}}
\global\def\subsubsection{\@startsection{subsubsection}{3}{\z@}%
                                     {-1ex\@plus -1ex \@minus -.1ex}%
                                     {0.1ex \@plus .1ex}%
                                     {\normalfont\itshape\fontsize{10}{12}\selectfont}}

\def\noeditingmarks{}
\newcommand{\textred}[1]{\textcolor{red}{#1}}
\ifx\noeditingmarks\undefined
   \newcommand{\pgwrapper}[2]{\textred{#1: #2}}
\else
   \newcommand{\pgwrapper}[2]{}
\fi

\usepackage{epsfig,comment}
\newcommand{\squishlist}{
   \begin{list}{$\bullet$}
    { \setlength{\itemsep}{0pt}      \setlength{\parsep}{3pt}
      \setlength{\topsep}{3pt}       \setlength{\partopsep}{0pt}
      \setlength{\leftmargin}{1.0em} \setlength{\labelwidth}{1em}
      \setlength{\labelsep}{0.5em} } }
\newcommand{\squishend}{
    \end{list}  }

\SetCommentSty{mycommfont}
\SetKwInput{KwInput}{Input}                %
\SetKwInput{KwOutput}{Output}              %

\usepackage{listings}

\usepackage[]{hyperref}
\usepackage{xurl}

\newcommand{\nop}[1]{}

\newcommand{\tightcaption}[1]{\vspace{-5pt}\caption{{\bf \small #1}}
\vspace{-15pt}
}
\newcommand{\para}[1]{\smallskip\noindent{\bf #1}}

\newcommand{\cut}[1]{}

\newcommand{\name}{Apparate\xspace}
\def\compactify{\itemsep=0pt \topsep=0pt \partopsep=0pt \parsep=0pt}
\let\latexusecounter=\usecounter

\AtBeginDocument{%
  }

\setcopyright{acmlicensed}

\acmJournal{JDS}
\acmVolume{37}
\acmNumber{4}
\acmArticle{111}
\acmMonth{8}

\usepackage{xspace}

\begin{document}

\title{\large \name: Rethinking Early Exits to Tame Latency-Throughput Tensions in ML Serving\vspace{-10pt}}

\def\refPrinceton{$^\mathsection$} %
\def\refGatech{$^\dagger$}

\author{\normalsize Yinwei Dai$^\star$\refPrinceton{}\hspace{0.5cm}Rui Pan$^\star$\refPrinceton{}\hspace{0.5cm}Anand Iyer\refGatech{}\hspace{0.5cm}Kai Li\refPrinceton{}\hspace{0.5cm}Ravi Netravali\refPrinceton{}\\}
{\vspace{0.2cm}
\author{\normalsize \refPrinceton{}Princeton University\hspace{0.5cm}\refGatech{}Georgia Institute of Technology}

\acmYear{2024}\copyrightyear{2024}
\setcopyright{rightsretained}
\acmConference[SOSP '24]{ACM SIGOPS 30th Symposium on Operating Systems Principles}{November 4--6, 2024}{Austin, TX, USA}
\acmBooktitle{ACM SIGOPS 30th Symposium on Operating Systems Principles (SOSP '24), November 4--6, 2024, Austin, TX, USA}
\acmDOI{10.1145/3694715.3695963}
\acmISBN{979-8-4007-1251-7/24/11}

\renewcommand{\shortauthors}{Trovato et al.}
\acmArticleType{Review}
\acmContributions{BT and GKMT designed the study; LT, VB, and AP
  conducted the experiments, BR, HC, CP and JS analyzed the results,
  JPK developed analytical predictions, all authors participated in
  writing the manuscript.}

\input{abstract}

\settopmatter{printfolios=true}
\settopmatter{printacmref=false}  %
\maketitle
\thispagestyle{empty}
\pagestyle{empty}

\blfootnote{$^\star$ Equal contributions.}
\begin{sloppypar}
\input{intro}

\input{background}

\input{design}

\input{eval}

\input{related}

\input{conclusion}

\label{EndOfPaper}
\end{sloppypar}

\label{lastpage}
\balance
\Urlmuskip=0mu plus 1mu\relax
\bibliographystyle{ACM-Reference-Format}
\bibliography{paper}
\newpage
\input{appendix}

\end{document}

%% file: abstract.tex
\begin{abstract}
Machine learning (ML) inference platforms are tasked with balancing two competing goals: ensuring high throughput given many requests, and delivering low-latency responses to support interactive applications. Unfortunately, existing platform knobs (e.g., batch sizes) fail to ease this fundamental tension, and instead only enable users to harshly trade off one property for the other. This paper explores an alternate strategy to taming throughput-latency tradeoffs by changing the granularity at which inference is performed. We present \name{}, a system that automatically applies and manages early exits (EEs) in ML models, whereby certain inputs can exit with results at intermediate layers. To cope with the time-varying overhead and accuracy challenges that EEs bring, \name{} repurposes exits to provide continual feedback that powers several novel runtime monitoring and adaptation strategies. \name{} lowers median response latencies by 40.5--91.5\% and 10.0--24.2\% for diverse CV and NLP classification workloads, and median time-per-token latencies by 22.6--77.9\% for generative scenarios, without affecting throughputs or violating tight accuracy constraints.%

\end{abstract}

%% file: intro.tex
\vspace{-15pt}
\section{Introduction}
\label{s:intro}

Machine Learning (ML) inference has become a staple for request handling in interactive applications such as traffic analytics, chatbots, and web services ~\cite{gemel,chatgpt,SpeechRecognition,web1}. To manage these ever-popular workloads, applications typically employ serving platforms~\cite{clockwork,tensorflowserving,onnx-rt,triton,infaas,clipper,vllm,orca} that ingest requests and schedule inference tasks with pre-trained models across large clusters of compute resources (typically GPUs). The overarching goals of serving platforms are to deliver sufficiently \emph{high throughput} to cope with large request volumes -- upwards of billions of requests per day~\cite{mlperf,ONNX1Trillion} --  while respecting the service level objectives (SLOs) that applications specify for response times (often 10--100s of ms). 

Unfortunately, in balancing these goals, serving platforms face a challenging tradeoff (\S\ref{ss:background}): requests must be batched for high resource efficiency (and thus throughput), but larger batch sizes inflate queuing delays (and thus per-request latencies). Existing platforms navigate this \emph{latency-throughput tension} by factoring only tail latencies into batching decisions and selecting max batch sizes that avoid SLO violations. Yet, this trivializes the latency sensitivity of many interactive applications whose metrics of interest (e.g., user retention~\cite{weblat1,weblat2}, safety in autonomous vehicles~\cite{autvehicles}) are also influenced by how far below SLOs their response times fall.

This paper explores the role that early exits (EEs) -- an adaptation mechanism that has garnered substantial ML research interest in recent years~\cite{branchynet,berxit,elbert,pabee,msdnet,deebert,fastbert} -- can play in resolving this tension. With EEs, intermediate model layers are augmented with ramps of computation that aim to predict final model responses. Ramp predictions with sufficiently high confidence (subject to a threshold) exit the model, foregoing downstream layers and bringing corresponding savings in both compute and latency. The intuition is that models are often overparameterized (especially with recent growth~\cite{visparams,DNNWorkloadAnalysisATC19}), and `easy' inputs may not require complete model processing for accurate results. Importantly, unlike existing platform knobs (e.g., batch size) that simply walk the steep latency-throughput tradeoff curve, EEs rethink the granularity of inference on a per-input basis. This provides a path towards lowering request latencies without harming platform throughputs. Indeed, across CV and NLP workloads, we find that optimal use of EEs brings 24.8--94.0\% improvement in median latencies for the same accuracy and throughput. %

Despite the potential benefits, EEs are plagued with practical challenges that have limited their impact (\S\ref{ss:challenges}). The primary issue is that EE proposals have solely come in the context of specific model architectures that impose fixed ramp designs and locations~\cite{deebert,branchynet,CALM,FREE}. The lack of guidance for integrating EEs into arbitrary models is limiting, especially given the ever-growing model offerings in the marketplace. Worse, even existing proposals lack any policy for \emph{runtime adaptation} of EE configurations, i.e., the set of active ramps and their thresholds. Such adaptation is crucial since dynamic workload characteristics govern the efficacy of each ramp in terms of exiting capabilities and added overheads (to non-exiting inputs); failure to continually adapt configurations can result in unacceptable accuracy drops up to 23.9\% for our workloads. However, devising adaptation policies is difficult: the space of configurations is massive, and it is unclear how to obtain a signal for accuracy monitoring once an input exits.

We present \textbf{\name{}}, the first system that automatically (i.e., without developer effort or expertise) injects and manages EEs for serving with a wide range of models. Our main insight is that the above challenges are not fundamental to EEs, and instead are a byproduct of what we are trying to get out of them. Specifically, adaptation challenges largely stem from halting execution for an input upon an exit, which leaves uncertainty in the `correct' response (as per the non-EE model). Instead, \name{} uses EEs \emph{only to deliver latency reductions}; results for successful exits are immediately released, but \emph{all} inputs continue to the end of the model. The key is in leveraging the (now) redundant computations to enable continual and efficient adaptation, while also remaining compatible with proven compute efficiency optimizations such as batching and model compression. %

Guided by this philosophy, \name{} runs directly atop existing serving platforms and begins by automatically converting registered models into EE variants. \name{}'s EE preparation strategy must strike a balance between supporting fine-grained runtime adaptation without burdening those time-sensitive algorithms with (likely) unfruitful options. To do so without developer effort, \name{} leans on guidance from the original model design, crafting ramp locations and architectures based on downstream model computations and data flow for intermediates around the model. Original model layers (and weights) are unchanged, and added ramps are rapidly trained in parallel (for efficiency), but in a manner that preserves their independence from other ramps.

Once deployed, \name{} continually monitors EE operation in GPUs, tracking computations and latency effects of each ramp, as well as outputs of the original model (for accuracy ground truth). To tackle the massive space of configuration options, \name{} judiciously decouples tunable EE knobs: thresholds for existing ramps are frequently and quickly tuned to ensure consistently high accuracy, while costlier changes to the set of active ramps occur only periodically as a means for latency optimization. For both control loops, \name{} leverages several fundamental properties of EEs to accelerate the tuning process. For instance, the monotonic nature of accuracy drops (and increases in latency savings) for higher thresholds motivates \name{}'s greedy algorithm for threshold tuning which runs 3 orders of magnitude faster than grid search while sacrificing only 0--3.8\% of the potential latency wins.

We evaluated \name{} across a variety of recent CV and NLP models (ranging from compressed to large language models), diverse workloads (classification and generative), and several serving platforms (TensorFlow-Serving~\cite{tensorflowserving}, Clockwork~\cite{clockwork}, and HuggingFace Pipelines~\cite{hf_pipelines}). Compared to serving without EEs, \name{} improves 25th percentile and median classification latencies by 70.2--94.2\% and 40.5--91.5\% for CV, and 16.0--37.3\% and 10.0--24.2\% for NLP, while imposing negligible impact on platform throughput. Latency wins are similar for generative scenarios: 22.6--77.9\% median time-per-token improvements. Importantly, unlike existing EE proposals that yield accuracy dips up to 23.9\%, we find that \name{}'s adaptation strategies \emph{always} met set accuracy constraints. We open source \name{} at https://github.com/dywsjtu/apparate.

%% file: background.tex
\section{Background and Motivation}
\label{s:background}

We start by overviewing existing ML serving platforms (\S\ref{ss:background}), highlighting the challenges they face in balancing metrics that are important for system performance (i.e., throughput, resource utilization) and application interactivity, i.e., per-request latencies. We then describe the promising role that early exits can play in alleviating those tensions (\S\ref{ss:ee}), and the challenges in realizing those benefits in practice (\S\ref{ss:challenges}). Results follow the methodology from \S\ref{ss:methodology}. %

\subsection{Model Serving Platforms}
\label{ss:background}

ML models are routinely used to service requests in interactive applications such as real-time video analytics~\cite{nexus,ekya}, chatbots~\cite{llama}, recommendation engines~\cite{ekko}, or speech assistants~\cite{preech}. To manage such workloads, especially at large scale, applications employ serving platforms such as ONNX runtime~\cite{onnx-rt}, TensorFlow-Serving~\cite{tensorflowserving}, PyTorch Serve~\cite{TorchServe}, Triton Inference Server~\cite{triton}, among others~\cite{orca,clockwork,clipper,nexus,infaas,vllm}. These platforms ingest pre-trained model(s), often in graph exchange formats like ONNX~\cite{onnx} and NNEF~\cite{nnef}, and are granted access to a pool of compute resources (usually with ML accelerators such as GPUs) for inference.

Given the latency-sensitive nature of interactive applications, requests are often accompanied with \emph{service level objectives} (SLOs) that indicate (un)acceptable response times for the service at hand. In particular, responses delivered after an SLO expires are typically discarded or yield severely degraded utility. Common SLOs are in the 10--100s of milliseconds, e.g., for live video analytics~\cite{nexus,gemel}.

During operation, serving platforms queue up incoming requests that can arrive at fixed or variable rates, and continually schedule jobs across the available compute resources. An inference task may be scheduled to run on a single node in a cluster, or may be distributed across multiple nodes~\cite{orca, clockwork}.

\begin{figure}[!tbp]
    \centering
    \includegraphics[width=\linewidth]{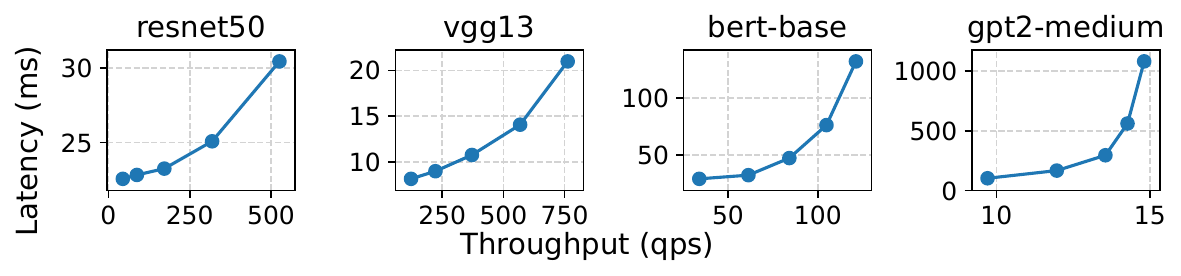}
    \vspace{-9pt}
    \tightcaption{Throughput-latency tradeoff in model serving. Results show serving times with batch sizes of 1--16.} 
    \vspace{1pt}
    \label{fig:tension}
\end{figure}

\para{Latency-Throughput tension.} To support the need for high \emph{throughput}, serving platforms resort to \emph{batching}, whereby inputs are grouped into a single high-dimensional tensor that moves through the model in lockstep, kernel by kernel, with final per-request responses being delivered at the same time. Larger batch sizes amortize the cost of loading a kernel into GPU memory across more inputs, and enable more effective use of accelerator parallelism~\cite{clipper,mark}. %

\begin{figure}[t]
    \centering
    \begin{subfigure}[b]{0.49\linewidth}
        \centering
        \includegraphics[width=\linewidth]{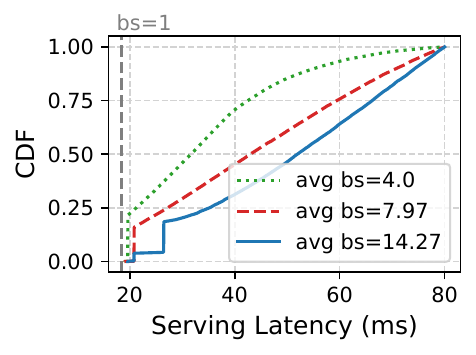}
                \vspace{-20pt}
        \caption{ResNet50}  
    \end{subfigure}
    \hfill
    \begin{subfigure}[b]{0.49\linewidth}  
        \centering 
        \includegraphics[width=\linewidth]{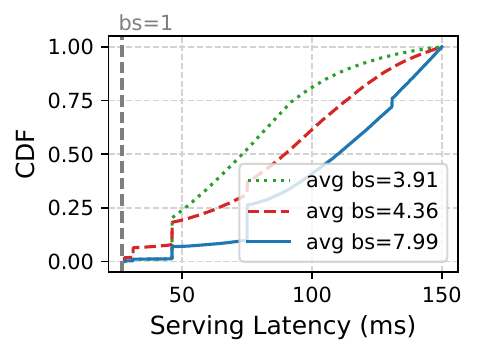}
        \vspace{-20pt}
        \caption{BERT-base}   
    \end{subfigure}
    \vspace{5pt}
    \tightcaption{Tuning platform knobs lowers latencies but harms throughput. Results vary TF-Serve's $max\_batch\_size$ from 4--16. Gray lines show min serving time per model (batch=1). CV uses a random corpus video; NLP uses Amazon reviews~\cite{amazon_reviews}.} 
    \label{fig:tuning}
\end{figure}

\begin{figure}[t]
    \centering
    \includegraphics[width=0.8\linewidth]{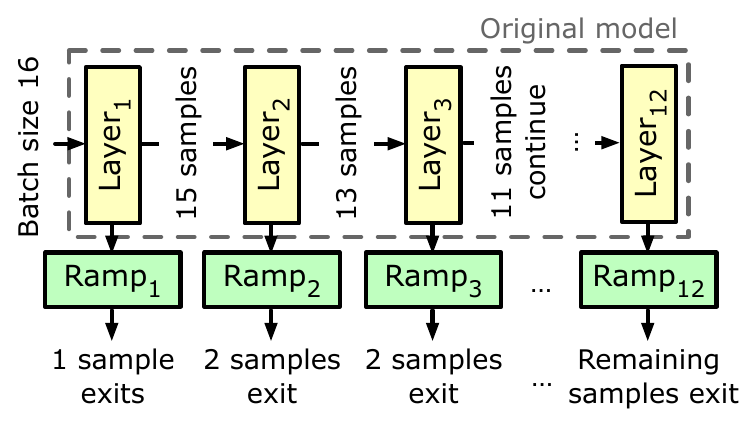}
    \tightcaption{Early exits enable termination of inputs at intermediate layers, lowering both compute and latency.}
    \vspace{6pt}
    \label{fig:ee_overview}
\end{figure}

Unfortunately, delivering the throughput necessary to support high request rates~\cite{mlperf,FacebookML} is directly at odds with \emph{per-request} latencies (Figure~\ref{fig:tension}). On one hand, latency for an input is minimized by scheduling inference as soon as the request arrives with batch size of 1. On the other hand, throughput is maximized by creating large batches using a queuing system which directly inflates request latencies.%

\para{The problem.} In navigating this tension, the key decision that serving platforms face is when to drain queued requests for inference. Certain platforms~\cite{clockwork,nexus,clipper} take an all-or-nothing stance on latency, with adherence to SLOs considered complete success, and violations viewed as failure. Accordingly, these platforms schedule inference jobs in a work-conserving manner and select the \emph{max} batch size that limits SLO violations for queued requests. However, many interactive applications present a more nuanced latency story where sub-SLO responses are not equally useful, e.g., faster responses boost conversational interactivity for chatbots~\cite{SpeechRecognition,speechlat} and confidence in scene perception for video analytics~\cite{odin,ekya}.

Other platforms~\cite{tensorflowserving,triton,TorchServe} provide more flexibility by exposing tunable knobs to guide queue management, e.g., $max\_batch\_size$ and $batch\_timeout\_micros$ parameters cap batch sizes or inter-job scheduling durations. However, such knobs do little to ease the throughput-latency tension, presenting harsh tradeoffs (Figure~\ref{fig:tuning}): tuning for median latency improvements of 17.3--39.1\% brings 1.1--3.6$\times$ reductions in average batch sizes (and proportional hits on throughput). 

Platforms for serving generative models~\cite{vllm,orca} face similar tensions despite the less explicit focus on SLOs (since sequence lengths are hard to predict). Indeed, although such platforms use continuous batching to ensure that new requests immediately leverage idle resources as any input's generation finishes, they prioritize throughput by running at the highest possible batch size (capped by a preset max).

\para{Takeaway.} Existing platform configurations and knobs fail to practically remediate the throughput-latency tension, and instead simply navigate (often) unacceptable tradeoff points between the two goals. Given ever-growing request rates and the need for high throughput, we ask if there is a middleground: whereby new serving adaptations enable lower per-request latencies (moving closer to the lower-bound serving times in Figure~\ref{fig:tuning}) without harming platform throughput.

\subsection{Early-Exit Models}
\label{ss:ee}

Early (or multi) exit models~\cite{branchynet, deebert} present an alternate way to address this tension by rethinking the granularity of inference. As shown in Figure~\ref{fig:ee_overview}, the key premise is that certain `easy' inputs may not require the full predictive power of a model to generate an accurate result. Instead, results for such inputs may be predictable from the values at intermediate layers. In such cases, the foregone model execution can yield proportional reductions in both per-request latencies and compute footprints. Thus, the goal with early exits (EEs) is to determine, on a per-input basis, the earliest model layer at which an accurate response can be generated.

To use EEs, intermediate layers in a model are augmented with \emph{ramps} of computation. These ramps ingest the values output by the layers they are attached to and parse them to predict the final model's result, e.g., a classification label. Ramps can perform arbitrary degrees of computation to arrive at a potential result. Exiting decisions at each ramp are made by comparing the entropy in the predicted result (or averaged over the past $k$ ramps) to a preset \emph{threshold}. Thresholds are set to balance latency and compute wins with potential dips in accuracy; a higher threshold implies lower required confidence for exiting, and thus more exiting.

\begin{figure}[t]
    \centering
    \begin{subfigure}[b]{0.49\linewidth}  
        \centering 
        \includegraphics[width=\linewidth]{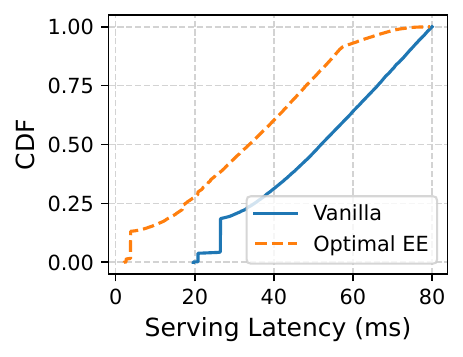}
        \vspace{-20pt}
        \caption{ResNet50}
    \end{subfigure}
    \hfill
    \begin{subfigure}[b]{0.49\linewidth}
        \centering
        \includegraphics[width=\linewidth]{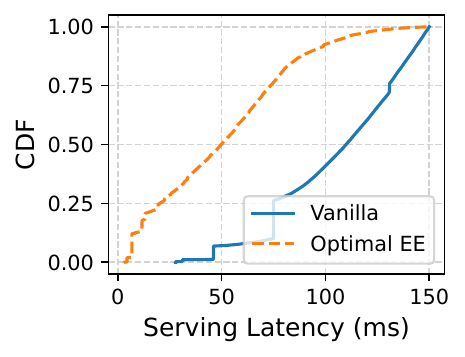}
                \vspace{-20pt}
        \caption{BERT-base}
    \end{subfigure}
    \vspace{3pt}
    \tightcaption{EEs can lower latencies without harming throughput. Results modulate latencies from TF-Serve with original/vanilla models (Figure~\ref{fig:tuning}) based on optimal exiting.}
    \vspace{2pt}
        
    \label{fig:ee_wins}
\end{figure}

\para{Potential benefits.} To understand the effect that EEs can have on the latency-throughput tension, we used off-the-shelf EE variants for the models in Figure~\ref{fig:tuning}: BranchyNet~\cite{branchynet} (CV) and DeeBERT~\cite{deebert} (NLP). For each model-input pair, we identified the \emph{optimal} exit point defined as the earliest exit ramp that predicted the correct response for the input. We then modified the highest-throughput results in Figure~\ref{fig:tuning} to account for exiting by subtracting the time saved for each exiting input, i.e., the difference in time for passing an input to the end of its optimal ramp versus passing it to the end of the model (without any ramps). These results are conservative upper bounds in that they do not reduce queuing delays or alter job scheduling. As shown in Figure~\ref{fig:ee_wins}, without changing queueing decisions, EEs can bring 35--54.7\% and 17.9--26\% improvements in median and 95th percentile latencies.%

\subsection{Challenges}
\label{ss:challenges}

Despite many EE proposals from the ML community~\cite{pabee,fastbert,deebert,msdnet,branchynet,berxit,RightTool,FREE,CALM}, and their potential benefits, multiple issues complicate EE use in practice, limiting adoption. %

\para{C1: Latency and resource overheads.} Although exiting can enable certain inputs to eschew downstream model computations, exit ramps impose two new overheads on serving. First, to be used, ramps must also be loaded into GPU memory which is an increasingly precious resource as models grow in size~\cite{orca,visparams,DNNWorkloadAnalysisATC19} and inference spreads to resource-constrained settings~\cite{gemel,mystify}. For instance, DeeBERT inflates memory requirements by 6.6\% compared to BERT-base. Second, certain inputs may be too ``hard'' to accurately exit at an intermediate ramp. In these cases, serving latency and throughput mildly worsen as unsuccessful exiting decisions are made, e.g., inputs that cannot exit at any ramp slow by 22.0\% and 19.5\% with BranchyNet and DeeBERT. %

\begin{figure}[t]
    \centering
    \begin{subfigure}[b]{\columnwidth}
        \centering
        \includegraphics[width=0.94\columnwidth]{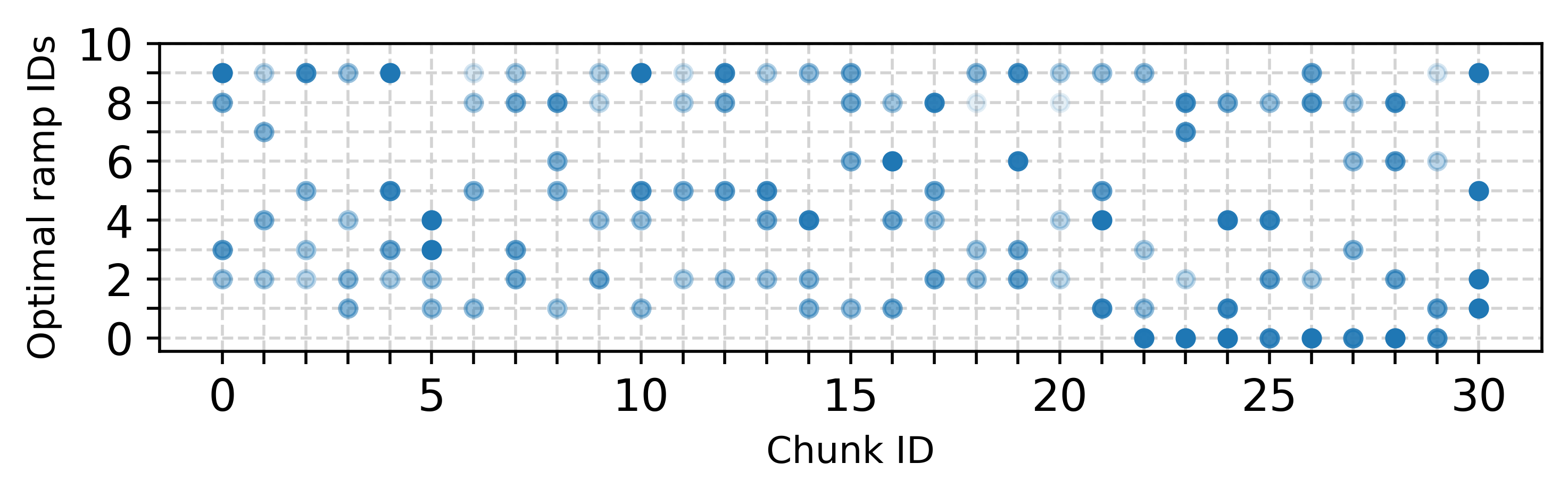}
        \vspace{-11pt}
        \caption{BERT-base; Amazon reviews}
        \vspace{-1pt}
    \end{subfigure}
    \hfill
    \begin{subfigure}[b]{\columnwidth}
        \centering
        \includegraphics[width=0.94\columnwidth]{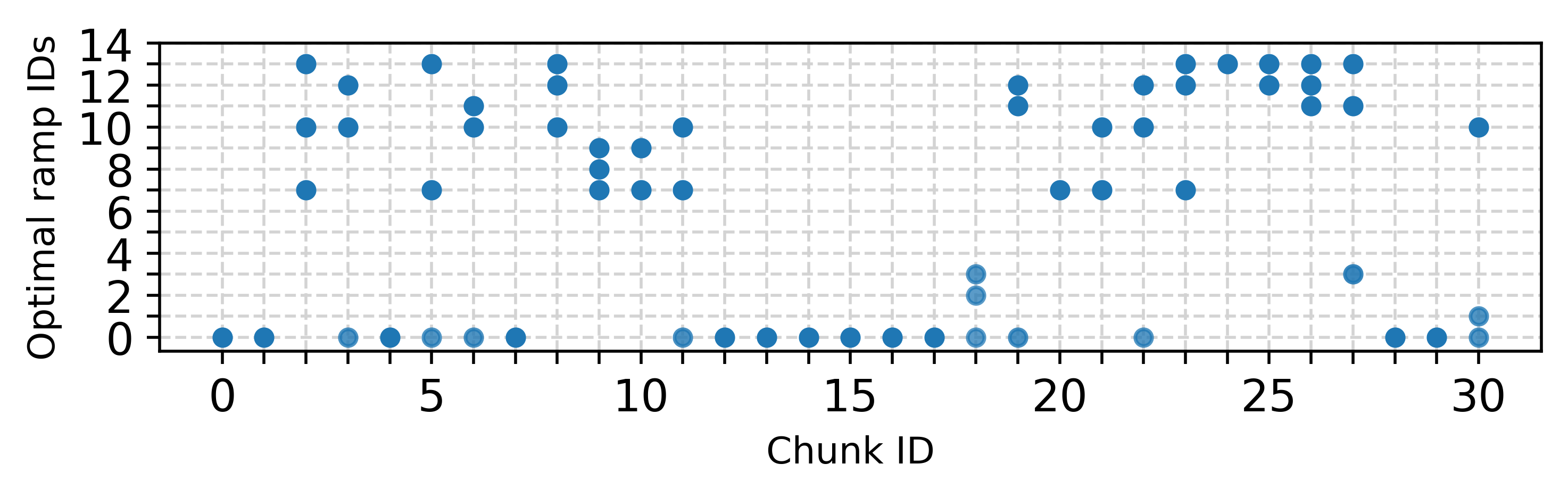}        \vspace{-11pt}
        \caption{ResNet50; random corpus video}
    \end{subfigure}
    \vspace{-5pt}
    \tightcaption{Optimal EE configurations change frequently. Workloads use 64-request chunks. Dot presence shows a ramp that was part of the optimal config for a chunk, while transparencies indicate threshold values (opaque is higher).}
    \vspace{6pt}
    \label{fig:opt_shifts}
\end{figure}

\begin{table}[t]
\footnotesize
  \centering
  \begin{tabular}{|c|c|c|}
    \hline
    \textbf{Strategy\textbackslash{}Workload} & \textbf{CV} & \textbf{NLP} \\
    \hline
    Initial Only & 84.5\% (74.3\%) & 86.8\% (73.6\%) \\
    \hline
    Uniformly Sampled & 90.3\% (64.2\%) & 87.7\% (69.4\%) \\
    \hline
    Continual Tuning & 98.6\% (43.5\%) & 98.3\% (26.6\%) \\
    \hline
  \end{tabular}
  \vspace{4pt}
  \tightcaption{Thresholds need frequent tuning to avoid accuracy loss. Continual tuning kicks in when chunk accuracy $<$ 99\%. Results list avg accuracy (median latency wins).}
  \label{t:cont_tuning}
\end{table}

\para{C2: Frequent and costly adaptation.} As shown in Figure~\ref{fig:opt_shifts}, the evolving nature of workloads for interactive applications brings frequent changes in the best EE configuration at any time, i.e., the set of active ramps (and their thresholds) that maximize latency savings without sacrificing response accuracy. Unfortunately, the large body of EE literature is unaccompanied by any policy for tuning ramps and thresholds during serving. Instead, proposed EE models are equipped with the max number of ramps, and mandate users to perform one-time tuning of thresholds. Such tuning is non-trivial and fails to cope with workload dynamism. For example, Table~\ref{t:cont_tuning} shows how one-time tuning on sampled data brings 8.3--14.5\% drops in accuracy relative to continual tuning. Worse, the space of configurations is untenably large, with many ramp options (i.e., at any layer, with any computation) and a continuous space of possible threshold values for each.

\para{C3: Lack of accuracy feedback.} %
EE ramp decisions are ultimately confidence-driven and may result in accuracy degradations (as shown above). In production scenarios, serving optimizations that deliver accuracy reductions $>$1--2\% are generally considered unacceptable~\cite{adaptivenet}. Yet, once deployed, EE models do not provide any indication of accuracy drops; indeed, when an exit is taken, the corresponding input does not pass through the remaining model layers, and the original (non-EE) model's prediction is never revealed. Thus, with early exiting, we lack mechanisms to determine when accuracy degradations are arising and EE tuning is required.

%% file: design.tex
\section{Design}
\label{s:design}

\name{} is an end-to-end system that automatically integrates early exits into models and manages their operation throughout the inference process. Its overarching goal is to optimize per-request latencies while adhering to tight accuracy constraints and throughput goals. Our key insight is in rethinking the way that EEs are configured and the benefits they are expected to deliver. In particular, rather than using EEs in the traditional way -- where inputs exit model inference to provide \emph{both} latency and computational benefits -- \name{} focuses solely on latency savings by allowing \emph{results} to exit, with inputs still running to completion. Foregoing true exiting (and thus, compute savings) grants \name{} with direct and continual feedback on EE accuracy (C3). This feedback provides the requisite signals for \name{} to continually adapt EE configurations to maximize latency savings while catering to resource constraints and workload dynamics (C1, C2).  %

\name{}'s design represents a departure from the typical expected utility of EEs (i.e., compute savings) that has been fraught with practical challenges. Instead, \name{} demonstrates an alternate avenue for benefits that EEs can bring (i.e., latency reductions), while remaining compatible with other compute efficiency optimizations that have had substantial practical traction. For instance, by foregoing true exiting, \name{} can run alongside request batching~\cite{griphook}. Further, \name{} supports diverse model architectures, including those that have been compressed for efficiency (\S\ref{ss:main_results}). We note that the redundant computations in \name{} match the work that vanilla serving perform by executing all model layers.

\begin{figure}[t]
    \centering
    \includegraphics[width=0.9\linewidth]{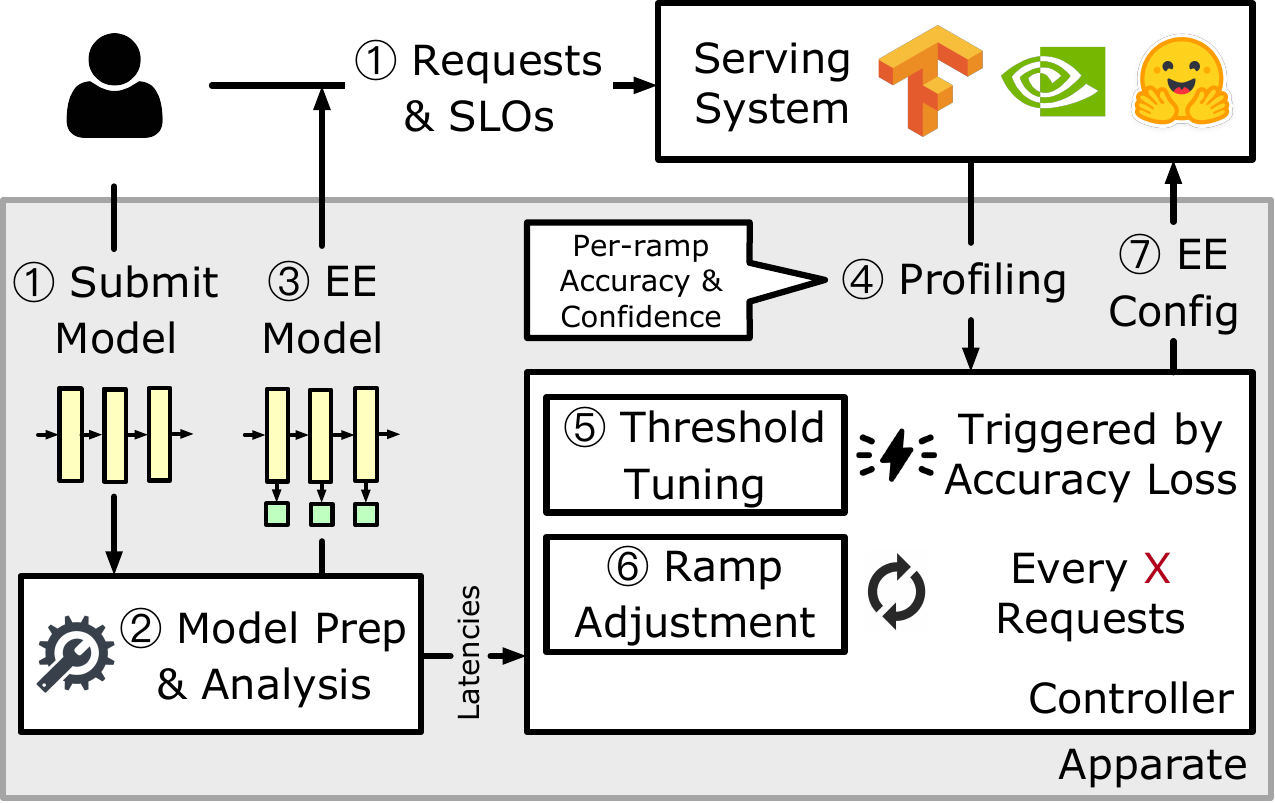}
    \vspace{6pt}
    \tightcaption{System architecture.}
    \label{fig:sys_arch}
\end{figure}

Figure~\ref{fig:sys_arch} overviews \name{}'s workflow, which runs atop existing serving platforms. Users register inference jobs as normal \numcircledtikz{1}, providing models and SLOs without needing any awareness of or expertise about EEs. In addition, \name{} introduces two parameters: (1) \emph{ramp aggression}, which bounds the number of active ramps in terms of \% impact on worst-case latency (and throughput), and (2) \emph{accuracy constraint} which indicates how much (if any) accuracy loss is acceptable relative to running the submitted model on all inputs without exiting. \name{}'s controller begins by configuring the provided model with EEs \numcircledtikz{2}, performing a graph assessment to determine suitable positions for ramps, and training those ramps on bootstrap data (\S\ref{ss:bootstrap}). The resulting model is passed to the serving platform for deployment \numcircledtikz{3}, after which \name{} shifts to management mode. %
In this phase, as requests arrive and inference is scheduled, 
\name{}'s controller gathers %
real-time feedback on the utility of each ramp (overheads vs. latency savings) and achieved accuracies (relative to the original model) \numcircledtikz{4}. This data is used to continually adapt the EE configuration \numcircledtikz{7} at different time scales: rapid threshold tuning for accuracy preservation (\S\ref{ss:threshold}) \numcircledtikz{5}, and less frequent ramp adjustments for latency optimization (\S\ref{ss:ramps}) \numcircledtikz{6}.

This decoupling of EE configuration adaptation into two tuning control loops is a key design decision that \name{} uses to manage the untenably large search space of ramp-threshold combinations (C1) without substantial loss in EE efficacy. Specifically, \name{} chooses to use frequent threshold tuning to preserve accuracy because it provides a finer-grained knob for walking the accuracy-latency tradeoff, i.e., thresholds are continuous, whereas ramp locations are inherently discretized. Thresholds also provide a mechanism to control ramp location; at the extreme, thresholds for any active ramps can be tuned to preclude exiting. Regardless, to limit foregone wins from infrequent ramp tuning, \name{} opts to employ many lightweight ramps (\S\ref{ss:bootstrap}): even if an optimal ramp is not present yet, a nearby ramp is likely active and can provide much of the same wins.

\para{Implementation details.} \name{} is implemented as a layer atop existing serving platforms (currently three~\cite{tensorflowserving,clockwork,hf_pipelines}, though its techniques generalize to others), and comprises $\sim$7500 lines of Python code for EE preparation (\S\ref{ss:bootstrap}) and management (\S\ref{ss:threshold}-\ref{ss:ramps}). \name{} runs a separate controller per model replica (as decided by the serving platform) on a CPU, with GPUs streaming per-ramp/batch profiling information in a \emph{non-blocking} fashion. This is possible since inputs pass to the end of models with \name{}, irrespective of exiting decisions. Tasks associated with model handling and serving are handled by the underlying serving platform, e.g., loading from disk, queuing, and batching.

\subsection{Preparing Models with Early Exits}
\label{ss:bootstrap}

Upon job registration with any DNN, \name{}'s initial task is to automatically prepare the model to leverage EEs without developer effort. This phase repeats any time the submitted model changes, e.g., continual retraining~\cite{ekya,odin,recl}. Note that, in the event that a developer provides an EE model, \name{} can forego any training and instead immediately begin managing its exit configurations (\S\ref{ss:threshold}--\ref{ss:ramps}). %

\para{Ramp locations.} \name{} accepts a model in the ONNX format, a widely used IR that represents the computation as a directed acyclic graph~\cite{onnx}. Once ingested, \name{} must first identify candidate layers for ramp addition. The goal is to maximize ramp coverage across the model (to provide more configuration options for \name{}'s runtime management), while avoiding ramps that are unlikely to be fruitful (but add complexity to adaptation decisions). To balance these aspects for diverse models, \name{} marks feasible ramp locations as those where operators are \emph{cut vertices}, i.e., a vertex whose removal would disconnect a graph into two or more disjoint sub-graphs. In other words, no edge can start before a ramp and re-enter the model's computation after the ramp.

\begin{figure}[t]
    \centering
    \includegraphics[width=0.93\linewidth]
    {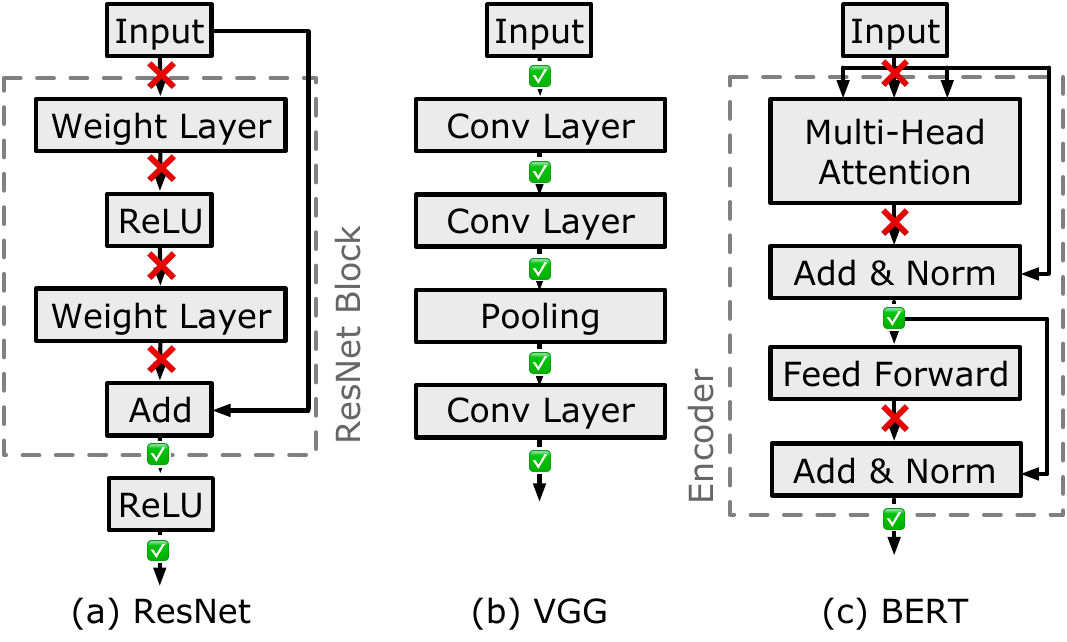}
    \vspace{4pt}
    \tightcaption{\name{} only injects ramps that make full use of available data flows at that part of the model.}
    \label{fig:cut_vertices}
\end{figure}

The idea is that such ramps take advantage of all available data outputs from the original model's processing to that point, boosting their chance at accurate predictions. As an example, consider families like ResNet or BERT 
which enable deep models by stitching together series of residual blocks, i.e., ResNet blocks for convolutions, or BERT encoders that each embed multi-head attention and feed-forward network residual blocks. To avoid performance degradations late in the model, the output of each block is ultimately a combination of its processing results and its input. In such scenarios, \name{} injects ramps between blocks, but not within each block to avoid ramps making decisions on partial data, i.e., ignoring block inputs. Ramps are similarly injected between trasnformer blocks of generative models, but only for decoding phases (as input tokens must be fully processed for generation). In contrast, for VGG models, ramps are feasible at all layers since their intermediates represent the full extent of data flow throughout the model. Figure~\ref{fig:cut_vertices} depicts examples. %

Overall, this strategy results in 9.2--68.4\% of layers having ramps for the models in our corpus, which we empirically observe is sufficient to adapt to dynamic workloads (\S\ref{ss:main_results}). However, we note that \name{} can directly support any other ramp configuration strategy, and offers a simple API for developers to express ramp policies or restrictions. %

\para{Ramp architectures.} For each feasible ramp location, \name{} must determine the style of ramp computations to use. Recall from \S\ref{s:background} that ramps can ultimately be composed of arbitrary layers and computations, with the only prerequisite being that the final layer sufficiently mimics that of the original model to ensure that response formats match. Determining the appropriate ramp complexity in this large space presents a tradeoff: additional computation can improve the exit capabilities of a ramp, but comes at the expense of (1) increased ramp latency, and (2) coarser flexibility and coverage at runtime since ramps become illogical if their computation exceeds that in the original model up until the next ramp.

\name{} opts for the shallowest ramps that can transform the intermediates at any layer into a final model prediction. Specifically, ramps comprise the model's final fully-connected (fc) layer, prepended with a lightweight pooling operation that reduces the dimensionality of intermediates to ensure compatability with the fc layer. This manifests differently for various model types. For instance, for vision models like ResNet, pooling is simply the model's penultimate layer. Similarly, for generative LLMs, ramps can simply use the final decoder head to transform intermediate hidden states. In contrast, for BERT, only the basic operator is drawn from the BERT pooler module, i.e., extracting the hidden state corresponding to the first token~\cite{bert}. For all models, the input width of the fc layer is modified to match the intermediates at each ramp location; the output remains unchanged to preserve result formats. %

\begin{figure}[t]
    \centering
    \includegraphics[width=0.7\linewidth]
    {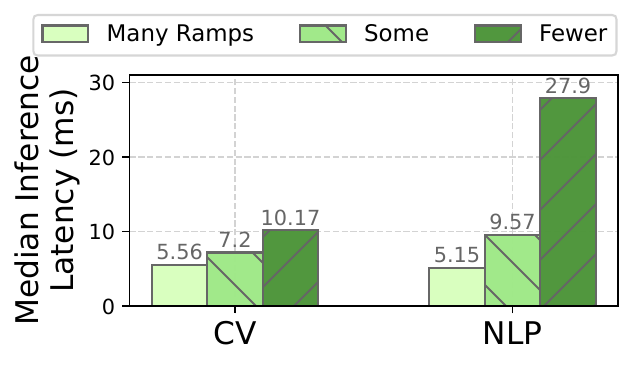}
    \tightcaption{More lightweight ramps boost EE savings. Results compare \name{}'s default `many ramps' with versions that use fewer, more expensive ramps.} %

    \label{fig:ramp_arch_justification}
\end{figure}

Figure~\ref{fig:ramp_arch_justification} evaluates this methodology by comparing with two, more expensive alternatives. With ResNet, to mimic model operations following each ramp, we add 1--2 convolution layers prior to pooling. For BERT, we consider two approaches: (1) add two fc layers after pooling, each with reduced width to shrink inputs to the final fc, and (2) following DeeBERT~\cite{deebert}, replace the simple pooling operator with the entire BERT pooler block and add a dropout as in the original model. In all cases, the number of ramps is subject to the same budget (i.e., \name{}'s default uses the most ramps), ramps are uniformly spaced across feasible positions in each model, and thresholds are optimally selected as in \S\ref{ss:ee}.

We observe that the added compute has minimal effect on ramp efficacy. For example, median latencies are 1.3--1.8$\times$ and 1.9--5.4$\times$ smaller with \name{}'s default ramps than the complex alternatives for CV and NLP. Nonetheless, to show \name{}'s generality, we consider other ramp styles in \S\ref{ss:deepdive}.

\para{Training ramps and deploying models.} To determine the appropriate weights for each ramp, \name{} can automatically label a dataset that is either developer-provided or collected online (running the vanilla model on live inputs). Automatic labeling is feasible since ramps aim to mimic the submitted model’s behavior (not ground truth), so the submitted model’s outputs can directly serve as labels. Regardless, during training, \name{} freezes the original model weights to ensure that non-EE behavior and feedback for tuning EEs is unchanged from the user's original intentions. Since its ramps are lightweight (single fully-connected layers) and comprise only 0.01--3.50\% of our models’ parameters, the FLOPs required for \name{}'s ramp training is significantly lower than whole-model pre-training or even fine-tuning. In cases where existing (final) layers can be used as ramps, e.g. in generative scenarios, \name{} eschews training and directly reuse the final layer for each ramp. %
In addition, \name{} enforces that \emph{all} inputs are used to train all ramps, i.e., exiting is prohibited during training. This ensures that ramps are trained \emph{independently} of the presence (or behavior) of any upstream ramps, which is crucial since the set of active ramps can vary at runtime. Further, such independence and model freezing enable loss calculations to be backwards propagated \emph{in parallel} across ramps, rapidly speeding up training despite the many lightweight ramps. As a result, ramp training only takes on the order of a few minutes for our models using a single A6000 GPU. \name{} uses the first 10\% of each dataset for training and validation (following a 1:9 split). %

For initial deployment, \name{} evenly spaces the max number of allowable ramps across the model. To avoid accuracy dips due to discrepancies between training data and the current workload, each ramp begins with a threshold of 0, i.e., no exiting. The updated model definition (with enabled ramps) is passed to the serving platform which runs normally.

\subsection{Accuracy-Aware Threshold Tuning}
\label{ss:threshold}

To avoid accuracy drops as workload characteristics change over time, \name{}'s controller employs frequent and fast tuning of thresholds for already-enabled ramps. The reason is that threshold tuning for any set of ramps is sufficient to ensure that user-specified accuracy constraints are not violated -- at the extreme, all thresholds could be set to zero, which precludes any early exiting. Altering only the set of active ramps fails to provide this property.

To enable threshold tuning, as requests pass through a model, \name{} continually records exiting information at each active ramp, as well as the final result that the original model predicts. More precisely, \name{} records the highest-confidence result for each ramp, even if the error exceeds the ramp's threshold (precluding exiting). Importantly, since inputs always pass fully through models with \name{}, this information is recorded for all inputs at each active ramp, irrespective of upstream exiting decisions. This is paramount since the information serves not only signals when to tune thresholds, but also provides guidance for how to do so.  %

\para{Triggering tuning.} \name{} maintains an average achieved accuracy over the past 16 samples by comparing exiting results with the deployed configuration to results of the original model. Threshold tuning is triggered any time a window's accuracy falls below the user-specified accuracy constraint. The threshold tuning process (described below) runs asynchronously on a CPU, without any disruptions to ongoing jobs. This is possible since thresholds are anyway enforced only by \name{}'s controller; GPUs are agnostic to threshold values, and instead simply stream ramp results to the \name{} controller which determines exiting decisions.

\para{Evaluating threshold configurations.} Threshold tuning needs insight into how any alterations to active ramp thresholds would affect model exiting behavior (and accuracies). By observing per-request behavior \emph{only at active ramps}, \name{} can rapidly evaluate any threshold configuration without additional inference, and while accounting for inter-ramp dependencies.
In particular, to evaluate new threshold values, \name{} simply identifies the earliest ramp whose top prediction now has an error rate below its threshold. Comparing these results with those of the original model indicates the achieved accuracy for the new configuration; latency wins are computed using the one-time profiling data described in \S\ref{ss:ramps}.

\para{Greedy search.} The goal of tuning is to identify a new set of thresholds that maximizes latency savings while adhering to accuracy constraints for the last window of data. The challenge is that the space of thresholds to consider is massive, precluding a grid search (especially given how frequently adaptation is needed - \S\ref{ss:challenges}). Indeed, even with discretized threshold values in [0, 1] with a step size of $S$, computation costs are $O(C \times (\frac{1}{S})^R)$, where $R$ is the number of active ramps, and $C$ is the cost to evaluate a given configuration.

\begin{figure}[!tbp]
    \centering
    \includegraphics[width=0.93\linewidth]{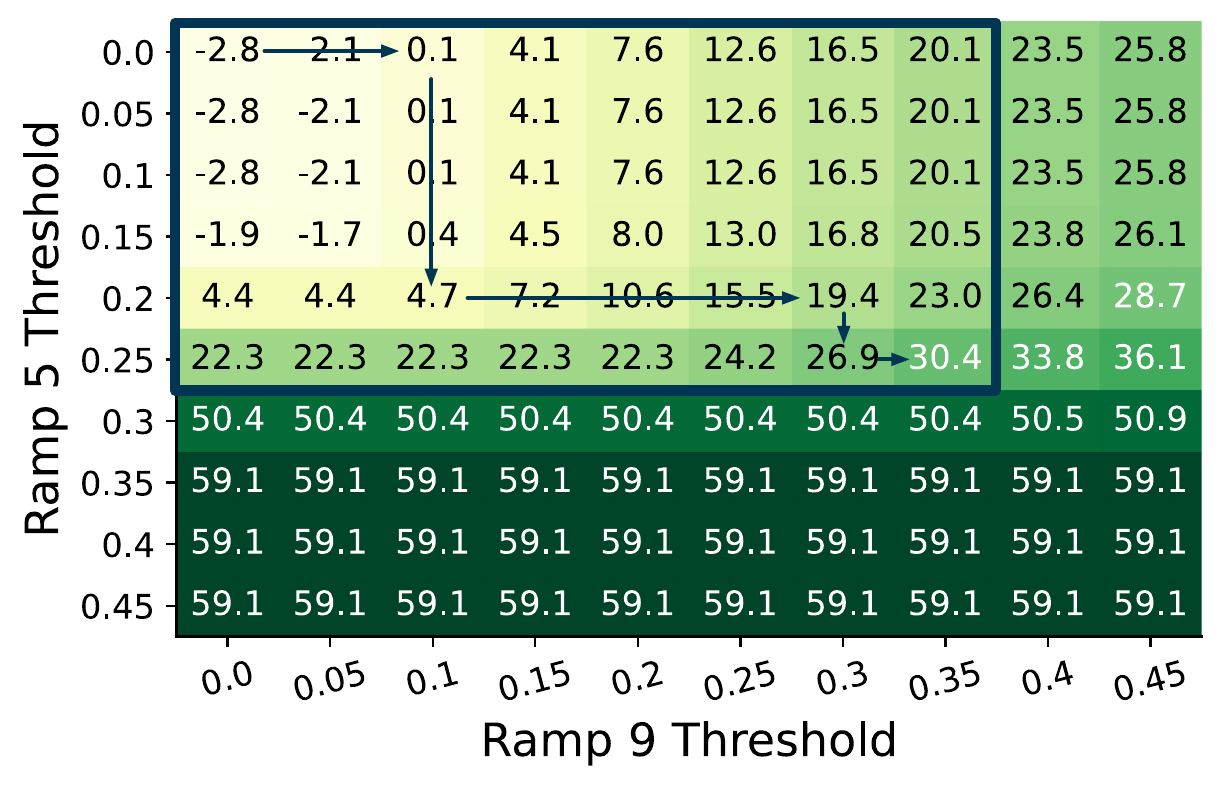}
    \tightcaption{Threshold tuning example with two active ramps for ResNet50 and a random video. Configurations within the boundary have $<$1\% accuracy loss; cell values list latency wins. Arrows show the path taken by \name{}'s hill climbing algorithm.}
    \vspace{1pt}
    
    \label{fig:threshold_tuning}
\end{figure}

Instead, \name{} employs a greedy heuristic that leverages a fundamental property of EEs when evaluated against an original model: \emph{higher thresholds result in monotonic decreases in accuracy and monotonic increases in latency savings}. This prunes the space of threshold values to consider by providing a clear boundary in the $R$-dimensional space that separates configurations that are sufficiently accurate from those that are not. Additionally, for accurate configurations, maximum latency savings must fall on that boundary. Figure~\ref{fig:threshold_tuning} illustrates this. %

These properties inform \name{}'s hill climbing strategy~\cite{hill_climbing} for threshold tuning. Starting with threshold values of 0 for each active ramp, and a step size of 0.1, threshold tuning runs in a series of (incremental) exploration rounds. In each round, we increase the threshold of each ramp in isolation (leaving the others unchanged), and evaluate the achieved accuracy and latency savings as described above. \name{} then chooses the single ramp threshold change that delivered the largest additional latency savings per unit of additional accuracy loss. This process repeats until no ramp's threshold can be increased without an accuracy violation. %

To enhance this process, \name{} follows a multiplicative increase, multiplicative decrease policy on step sizes to balance search speed and granularity. Each time a step increase results in an accuracy violation, \name{} halves that ramp's step size for subsequent rounds to hone in on the boundary at fine granularity; step sizes are lower-bounded at 0.01. Conversely, selection of a ramp for threshold alteration suggests a promising path of exploration; in this case, for a speedup, \name{} doubles that ramp's step size for the following round.

\begin{figure}[t]
    \centering
    \begin{subfigure}[b]{0.5\linewidth}  
        \centering 
        \includegraphics[width=\linewidth]{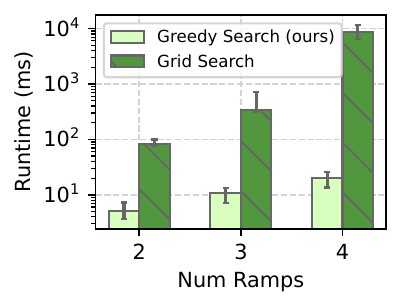}
        \vspace{-22pt}
        \caption{Threshold tuning speed.}    
    \end{subfigure}
    \hfill
    \begin{subfigure}[b]{0.47\linewidth}
        \centering
        \includegraphics[width=\linewidth]{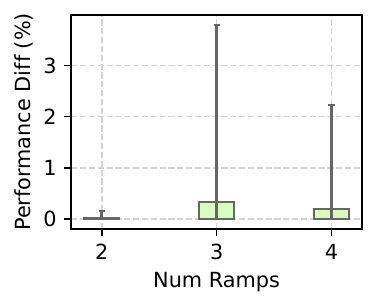}
                \vspace{-22pt}
        \caption{Optimality of tuning results.}
    \end{subfigure}
    \vspace{5pt}
    \tightcaption{\name{}'s tuning vs. optimal tuning on runtime and latency of selected configurations. Bars list medians across all model-workload pairs, with error bars for min-max.}
    \label{fig:tt_performance}
\end{figure}

Overall, as shown in Figure~\ref{fig:tt_performance}, \name{}'s threshold tuning algorithm runs up to 3 orders of magnitude faster than a pure grid search (11.9ms vs. 3.0s on average). %
Note that these results maximally parallelize grid search across a 32-core machine. Further, selected threshold values achieve within 0--3.8\% of the latency savings of the optimal configurations. %

\subsection{Latency-Focused Ramp Adjustments}
\label{ss:ramps}

The set of active ramps ultimately dictates where inputs can exit, and thus provides bounds on potential latency savings. Unlike threshold tuning which runs reactively (since accuracy is a constraint) and uses only recent profiling data to evaluate new configurations, ramp adjustment is used strictly as an optimization (for latency savings) in \name{}, and requires deployment to evaluate the impact of any new ramp. 
Thus, \name{}'s ramp tuning runs periodically (every 128 samples by default) and conservatively alters the set of active ramps to incrementally converge on high-performing configurations.

\para{Evaluating active ramps.} In each round, \name{}'s controller starts by computing a utility score for each active ramp that evaluates its overall impact on workload latency. To do so, \name{} couples per-ramp exit rates (\S\ref{ss:threshold}) with two additional inputs that are collected once per model during bootstrapping: (1) the latency overhead per ramp, and (2) a layer-wise breakdown of time spent during model inference (for different batch sizes~\cite{clockwork} and including network delays for distributed serving). The latter is necessary since latency characteristics vary across models but govern the impact of exits, e.g., latency arises early in CV models~\cite{gemel}, but more evenly across coding blocks in transformers. Using these inputs, \name{} defines the utility of ramp $R$ as \texttt{savings - overheads}, where \texttt{savings} is the sum of raw latency that exiting inputs avoided by using ramp $R$, and \texttt{overheads} is the sum of latency that $R$ added to inputs that it could not exit.%

\para{Adding new ramps.}  If any negative utility values exist, \name{} applies a fast threshold tuning round to see if ramp utilities become entirely positive without harming overall latency savings. If not, \name{} immediately deactivates all negative-utility ramps. From there, the key question to address is what ramps (if any) should be added to make use of the freed ramp budget. The main difficulty is in predicting the utility of each potential addition. Indeed, while per-exit latency savings for each potential ramp are known (using the latency breakdown from above), exit rates are not.

To cope with this uncertainty, our guiding intuition is that, subject to the same accuracy constraint, \emph{later ramps almost always exhibit higher exit rates than earlier ones}. %
The reason is that late ramps have the luxury of leveraging more of an original model's computations when making a prediction. Importantly, this implies that a candidate ramp's exit rate is bound by the exit rate of the closest downstream ramp; note that this is not a formal guarantee~\cite{overthinking}, and \name{} uses this for search efficiency (not correctness).

Building on this, \name{}'s controller computes an upper bound on the utility of candidate ramps as follows. To avoid inter-ramp dependencies harming ramps that are already performing well, we only consider additions after the latest positive ramp $P$ in the model. In particular, \name{} divides the range following $P$ into intervals separated by any negative ramps deactivated in this round. The first round of candidate ramps are those in the middle of each interval.

\begin{figure}[t]
    \centering
    \includegraphics[width=0.95\linewidth]{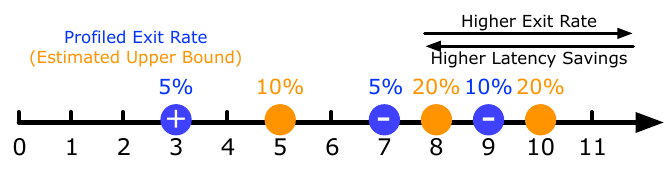}
    \tightcaption{Computing upper-bound exit rates for candidate ramps. Blue dots show previously active ramps ($+/-$ indicates positive/negative utility), while orange dots show candidates.}
    \vspace{1pt}%
       
    \label{fig:two_ramps}
\end{figure}

For each candidate ramp, we compute its upper-bound exit rate as the sum of profiled exit rates for the following deactivated ramp and any earlier deactivations (Figure~\ref{fig:two_ramps}); the idea is that inputs from earlier deactivations would have reached the following deactivated ramp and \emph{might} have exited there. Utility scores are then computed as above, and the ramp with the highest positive utility score is selected for trial. If all ramps have negative projected utilities, \name{} repeats this process for later candidate ramps in each interval. Once a ramp is selected for trial, \name{} adds it to the deployed model definition, while removing deactivated ramps. Trialed ramps start with threshold=0 to prevent inaccurate exiting, but are soon updated in the next round of threshold tuning.

Until now, we have only discussed how \name{} handles scenarios with at least one negative ramp utility. In the event that all ramps exhibit positive utilities, \name{} enters a low-risk probing phase to determine if latency savings can grow by using earlier ramps. If ramp budget remains, we add a ramp immediately before the existing ramp with \emph{highest} utility (while keeping that ramp to preserve its exiting wins). If not, we shift the ramp with the \emph{lowest} utility score one position earlier, leaving the most positive ramp untouched. %

\subsection{Supporting Generative Large Language Models}
\label{ss:ee_llm}

Recent efforts have incorporated EEs into generative language models (e.g., CALM~\cite{CALM} and FREE~\cite{FREE} for T5~\cite{t5}, LayerSkip~\cite{layerskip} for Llama~\cite{llama}) to enhance their response times in interactive applications. While the general challenges with EEs persist -- lack of runtime adaptation for time-varying accuracy and latency savings -- the auto-regressive nature of generative models poses a unique challenge for \name{}. Unlike with classification where inputs can be processed to completion \emph{independently}, each token generated by an LLM depends on the preceding ones in the sequence, requiring their full key-value (KV) states. 
Thus, when a token $T1$ exits using a ramp, the generation delay begins to accumulate for the next token, $T2$; yet, $T2$'s forward pass through the layers after the ramp cannot begin until $T1$ passes through the same layers to generate KV states. The effect is potential harm on time-per-token (TPT) distributions.

To regain latency savings from EEs, \name{} draws inspiration from recent parallel decoding techniques~\cite{FREE,specdecode,layerskip}. 
As a token exits from a ramp, \name{} \emph{does not immediately compute the remaining model layers}, and instead only accumulates its hidden states at that ramp.
The remaining computations are executed in parallel alongside the first non-exiting token that is encountered thereafter at that ramp. For concreteness, consider a scenario with 2 tokens, $T1$ and $T2$. Assume $T1$ can exit at ramp $R1$, while $T2$ must proceed later in the model. $T1$ will exit at $R1$, and $T2$ will immediately begin processing. However, once $T2$ fails to exit at $R1$, the remaining layers for $T1$ are run alongside (i.e., batched with) $T2$’s remaining layers. Taken together, $T1$ achieves per-token latency savings from exiting, while $T2$ incurs a very mild (\S\ref{ss:sota_llm}) penalty from the higher batch size. %

In addition to improved TPT latencies and compute efficiency (due to batching effects on GPUs), parallel decoding grants \name{} token-level feedback for exiting decisions relative to the original (non-EE) model. Specifically, for each parallel decoding instance, \name{} collects per-token feedback up until the first token whose exit result deviates from the original model; feedback for subsequent tokens is discarded as it may reflect cascading errors from inter-token dependencies. This feedback guides \name{}'s threshold and ramp tuning strategies from \S\ref{ss:threshold}--\ref{ss:ramps}.

%% file: eval.tex
\section{Evaluation}
\label{s:eval}

We evaluated \name{} across a wide range of NLP and CV workloads and serving platforms. Our key findings are:

\squishlist

\item \name{} lowers 25th percentile and median classification latencies by 40.5--91.5\% and 70.2--94.2\% for CV, and 16.0--37.3\% and 10.0--24.2\% for NLP workloads, compared to original (non-EE) models. These wins are 5.7--66.6\% larger than two-layer inference systems using compressed models for `easy' inputs~\cite{tabi,canel2019scaling}. Median time-per-token wins are 22.6--77.9\% for generative scenarios.

\item Unlike existing EE models that unacceptably worsen accuracies and tail latencies by up to 23.9\% and 11.0\% for classification, \name{} consistently meets accuracy and tail latency constraints. This carries to generative scenarios: \name{}'s tail latency is 1.8--2.4\% lower than existing EE models which lower accuracies up to 5.5\%.

\item \name{} automatically generalizes to different model architectures (e.g., compressed) and EE configurations (e.g., ramp style), and its wins gracefully shrink as accuracy or tail-latency constraints grow.

\squishend

\subsection{Methodology}
\label{ss:methodology}

\para{Models.} For \emph{classification}, we consider 10 models (across 4 families) that cover popular architectures and diverse sizes. For CV, we use the ResNet\{18, 50, 101\} residual models, as well VGG\{11, 13, 16\} models that follow a chained (linear) design. All of these models are pre-trained on ImageNet and from the PyTorch Model Zoo~\cite{torchzoo}. For NLP, we consider 3 encoder-only transformers from the BERT family -- BERT-base, BERT-large, and Distilbert~\cite{distilbert} (a variant of BERT-base that was compressed via distillation) -- as well as a decoder-only LLM: GPT2-medium. These models span 66--345 million parameters, were collected from HuggingFace~\cite{hface}, and were pre-trained on Yelp reviews~\cite{yelp_dataset}. We also consider quantized versions of these BERT models in \S\ref{ss:main_results}. For \emph{generative scenarios}, we use the T5-large~\cite{t5} encoder-decoder LLM (from prior EE work~\cite{CALM,FREE}) 
with 770 million parameters and the Llama2 decoder-only LLMs with 7 and 13 billion parameters, both pretrained from HuggingFace. 

\para{Workloads.} CV workloads comprise real-time object classification (people, cars) on 8 one-hour videos from recent video analytics literature~\cite{boggart,focus}. The videos were sampled at 30 frames per second, and span day/night from urban scenes.%

NLP classification workloads focus on sentiment analysis using two datasets: Amazon product reviews~\cite{amazon_reviews} and IMDB movie reviews~\cite{imdb}. To the best of our knowledge, there do not exist public streaming workloads for NLP classification, so we convert these datasets into streaming workloads as follows. For Amazon, we follow the order of product categories in the original dataset, but within each category, we keep reviews only from frequent users (i.e., those with $>$1k reviews) and order streaming by user (250k requests in total). For IMDB, we follow the order of reviews in the original dataset, but stream each in sentence by sentence (180k requests in total). We then define arrival patterns using the Microsoft Azure Functions (MAF) as in prior work~\cite{clockwork,alpaserve}. To cope with the large variation in runtime across our models, we paired each model with a randomly selected trace snippet from the set that met the following criteria: (1) number of requests match that in our largest dataset, and (2) queries per second should not result in $>$20\% dropped requests with vanilla serving for the given model and selected SLO (described below).

Our generative workloads use two datasets: CNN/DailyMail~\cite{cnn_dm} for text summarization and SQuAD~\cite{squad} for question answering. As in prior work~\cite{CALM,FREE}, we assign request arrival times that follow a Poisson distribution, configured to saturate our computing resources.%

\para{Parameter configurations.} Given our focus on interactivity, we cope with heterogeneity in model runtimes by setting SLOs for classification to be 2$\times$ each model's inference time with batch size 1 in our main experiments. This results in SLOs between $\sim$10--200 ms, which match the ranges used in prior work~\cite{nexus,gemel,clockwork}; Table~\ref{t:slos} in \S\ref{s:appendix} lists the specific SLO values, and we study the effect of SLO on \name{} in \S\ref{ss:deepdive}. Unless otherwise noted, results use 1\% for \name{}'s accuracy constraint and a ramp budget of 2\% impact on worst-case latency; we consider other parameter values in \S\ref{ss:deepdive}.

\para{Setup.} Experiments were conducted on dedicated servers with NVIDIA RTX A6000 GPUs housing 48GB of memory, AMD EPYC 7543P 32-Core CPUs, and 256GB DDR4 RAM. We run experiments with two serving platforms for classification: TensorFlow-Serving~\cite{tensorflowserving} and Clockwork~\cite{clockwork}. For space, we primarily present results with Clockwork, but note that reported trends hold for both platforms; we compare cross-platform results in \S\ref{ss:deepdive}. For generative workloads, we use the HuggingFace Pipelines inference engine~\cite{hf_pipelines}.

\para{Metrics and baselines.} For classification, our main metrics are accuracy (\% of inputs assigned correct label as per the non-EE model) and per-request response latency (including queuing). For generative models, sequence accuracy is measured using ROUGE-L (text summarization) and F1 (question answering) scores. Latency is measured using time-per-token (TPT) distributions. We compare \name{} with the following baselines: (1) 
original models without EEs (\emph{vanilla}), (2) \emph{optimal} EEs as defined in \S\ref{ss:ee}, i.e., assuming all inputs exit at their earliest possible ramps, with no ramp overheads, (3) two-layer inference systems~\cite{canel2019scaling,tabi} that invoke full models only when compressed ones cannot generate high-confidence outputs, and (4) existing, non-adaptive EE strategies (\S\ref{ss:sota}).

\begin{figure}[t]
    \centering
    \includegraphics[width=\linewidth]{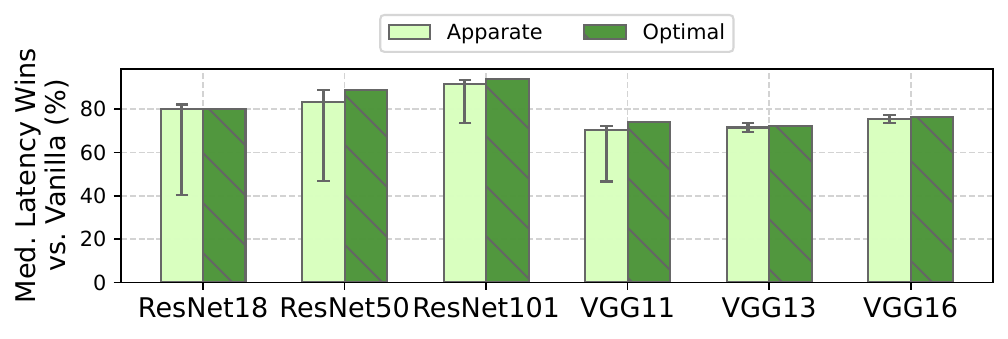}
    \vspace{-6pt}
    \tightcaption{Median latency savings vs. vanilla models. Bars show median savings across all workloads; error bars are min-max.}    
    \label{fig:cv_med}
\end{figure}

\begin{figure}[t]
    \centering
    \includegraphics[width=\linewidth]{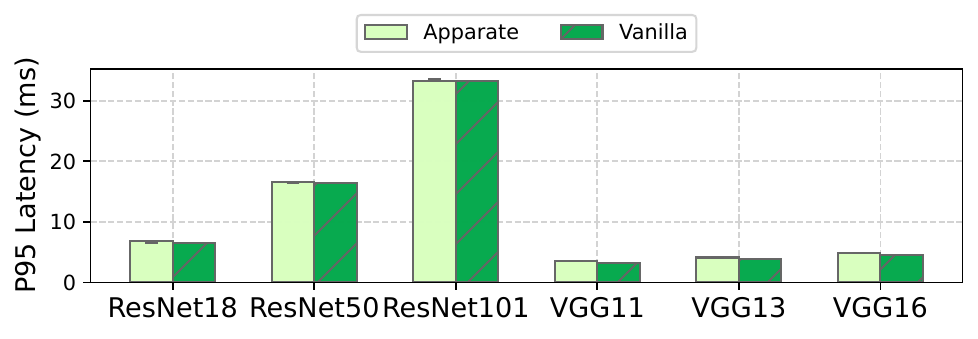}
    \vspace{-8pt}
    \tightcaption{Evaluating \name{}'s impact on tail latency (running with 2\% budget) vs. vanilla serving. Bars show the median savings across all workloads with error bars spanning min-max.}   
    \label{fig:cv_tail}
\end{figure}

\begin{figure*}
    \centering
    \begin{subfigure}{0.245\textwidth}
        \centering
        \includegraphics[width=\textwidth]{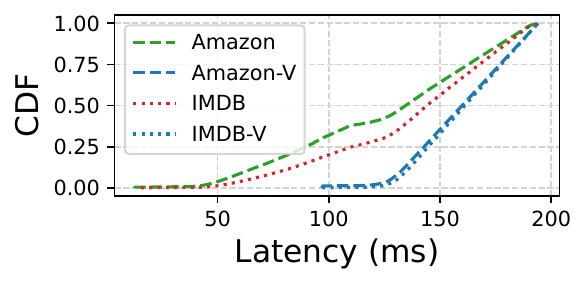}
        \vspace{-20pt}
        \caption{GPT2}
    \end{subfigure}
    \begin{subfigure}{0.245\textwidth}
        \centering
        \includegraphics[width=\textwidth]{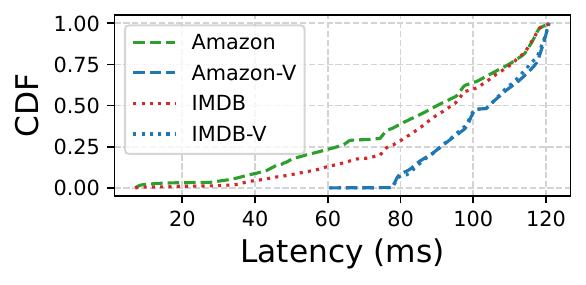}
                \vspace{-20pt}
        \caption{BERT-large}
    \end{subfigure}
    \begin{subfigure}{0.245\textwidth}
        \centering
        \includegraphics[width=\textwidth]{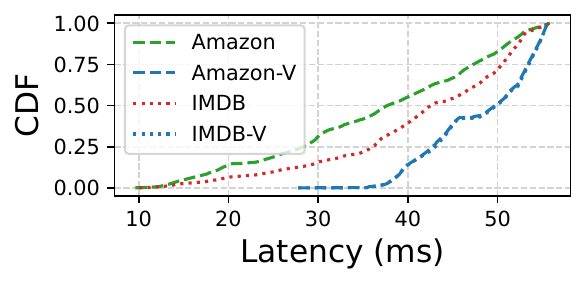}
                \vspace{-20pt}
        \caption{BERT-base}
    \end{subfigure}
    \begin{subfigure}{0.245\textwidth}
        \centering
        \includegraphics[width=\textwidth]{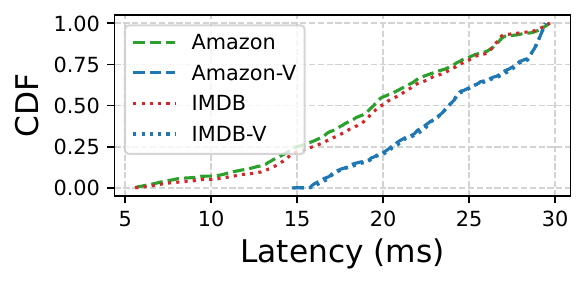}
                \vspace{-20pt}
        \caption{Distilbert-base}
    \end{subfigure}
    \vspace{6pt}
    \tightcaption{\name{} (with 2\% budget) vs. vanilla models (``-V'') on NLP classification workloads. ``-V'' curves per plot mostly overlap since they use the same timing trace and no exiting; minor discrepancies are due to the varying number of inputs across workloads.}%
    \vspace{1pt}
    \label{fig:nlp_cdf}
\end{figure*}

\begin{figure}[!tbp]
    \centering
    \includegraphics[width=0.7\linewidth]{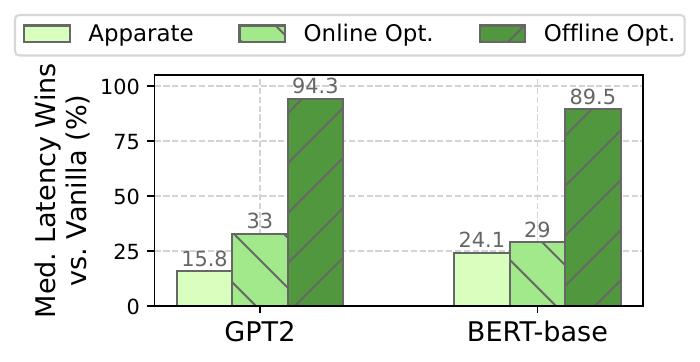}
    \tightcaption{\name{} vs. optimal exiting on NLP classification workloads with the Amazon dataset.}  
    \label{fig:optimal_comparison}
\end{figure}

\subsection{Results for CV and NLP Classification}
\label{ss:main_results}

Figures~\ref{fig:cv_med}--\ref{fig:optimal_comparison} compare \name{} with vanilla model serving and optimal exiting across our workloads. Overall, \name{} significantly lowers latencies compared to serving vanilla models, while always adhering to the imposed 1\% accuracy constraint. For instance, median speedups range from 40.5--91.5\% (2.7--30.5 ms) for CV workloads, and jump to 70.2--94.2\% (5.2--31.4 ms) at the 25th percentiles. NLP workloads follow a similar pattern, with median and 25th percentile savings ranging from 10.0--24.2\% (3.9--25.3 ms) and 16.0--37.3\% (4.8--53.2 ms). Importantly, across all workloads, \name{}'s tail latency (and thus impact on throughput) always falls under its 2\% budget, and is most often negligible (Figure~\ref{fig:cv_tail}). %

Beyond this, there are several important trends to note. First, \name{}'s raw latency savings grow with increased model sizes, e.g., 25th percentile wins of 53.2ms, 28.4ms, 14.3ms, 5.5ms for GPT-2, BERT-large, BERT-base, and Distilbert-base on the Amazon Reviews. This is because only results (not inputs) exit models with \name{}, so latency savings pertain entirely to serving times (not queuing delays) and grow as model size/runtime grows.
Relative (\%) latency savings follow the same pattern for CV workloads, e.g., \name{}'s median wins grow by 13.8\% and 5.3\% moving from the smallest to the biggest models in the ResNet and VGG families. However, relative wins remain relatively stable in NLP models, e.g., 15.8\% and 13.7\% for GPT2 and BERT-large on Amazon Reviews. The difference is due to the effectiveness of the models in each domain. Results and task performance are largely similar across the CV models, enabling \name{} to inject ramps early in (even larger) models. In contrast, results are far better with the larger models in NLP; thus, \name{}'s ramps fall in similar (relative) positions across the models. %

Second, \name{}'s wins are larger for CV workloads than NLP workloads for two reasons. As previously noted, CV workloads use lighter models and lower request rates (bound by video fps), and thus incur far lower queuing delays. More importantly, in contrast to CV where spatiotemporal similarities across frames (and thus, requests) are high due to physical constraints of object motion in a scene, NLP requests exhibit less continuity, e.g., back-to-back reviews are not constrained in semantic similarity. The effects on \name{}'s adaptations are that (1) past data is less representative of future data, and (2) the duration until subsequent adaptation is shorter.

\para{Other compute optimizations.} To further illustrate \name{}'s ability to run alongside existing compute efficiency optimizations -- a key goal of its design (\S\ref{s:design}) -- we ran experiments using post-training Int8 quantized Bert-base and Bert-large. Overall, we observe that \name{}'s speedups largely persist, which median and 25th percentile wins of 7.3--19.4\% and 6.9--31.1\%. The mild dip in speedups relative to non-quantized Bert models (Figure~\ref{fig:nlp_cdf}) is a result of quantization's reduction in model overparameterization, which early exiting aims to capitalize on for select inputs.

\begin{figure}[t]
    \centering
    \begin{subfigure}[b]{0.45\linewidth}  
        \centering 
        \includegraphics[width=\linewidth]{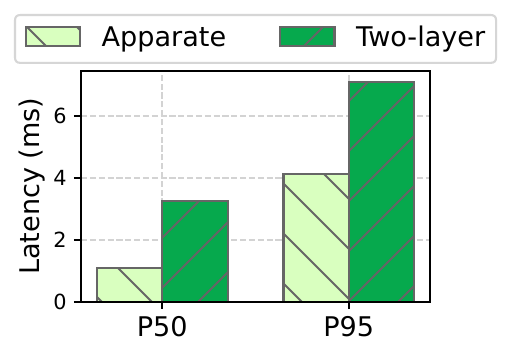}
        \vspace{-22pt}
        \caption{CV: VGG11, VGG13.}    
    \end{subfigure}
    \hfill
    \begin{subfigure}[b]{0.54\linewidth}
        \centering
        \includegraphics[width=\linewidth]{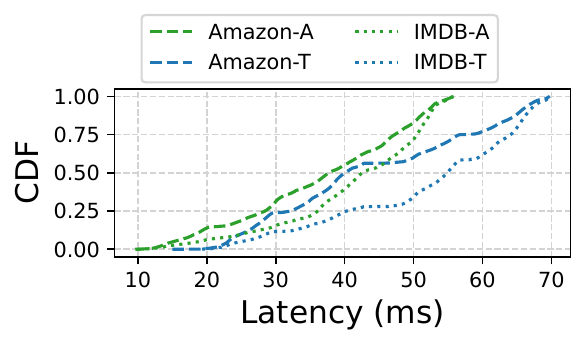}
                \vspace{-22pt}
        \caption{NLP: Distilbert, Bert-base}
    \end{subfigure}
    \vspace{-2pt}
    \tightcaption{\name{} (``-A'') vs. two-layer inference systems (``-T'').}  %
    \label{fig:baseline}
\end{figure}

\begin{figure}[t]
    \centering
    \begin{subfigure}[b]{0.49\linewidth}  
        \centering 
        \includegraphics[width=\linewidth]{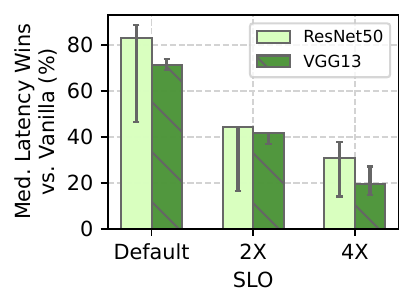}
        \vspace{-22pt}
    \end{subfigure}
    \hfill
    \begin{subfigure}[b]{0.49\linewidth}
        \centering
        \includegraphics[width=\linewidth]{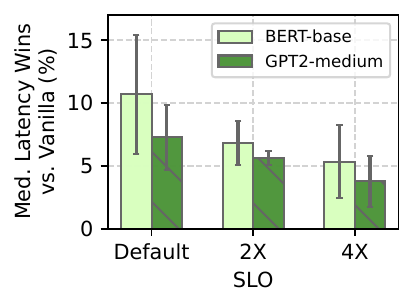}
                \vspace{-22pt}
    \end{subfigure}
    \vspace{10pt}
    \tightcaption{Impact of SLOs on \name{}'s wins.} %
    
    \vspace{2pt}
    \label{fig:slo_sweep}
\end{figure}

\para{Comparisons with optimal.} As shown in Figure~\ref{fig:cv_med}, latency savings with \name{} for CV workloads largely mirror those of the optimal that tunes exiting decisions based on perfect knowledge of the upcoming workload, e.g., median savings are within 20.5\% of the optimal. In contrast, the limited continuity across inputs in NLP workloads leads to a wider gap of 65.4--78.5\% at the median (Figure~\ref{fig:optimal_comparison}). To further characterize \name{}'s performance on these workloads, we also consider a more realistic \emph{online optimal} algorithm that relaxes the following elements. First, rather than per-sample adaptation of thresholds and ramps, ramp adjustments are set to operate only as fast as model definitions in the GPU can be updated. Second, rather than using perfect knowledge of upcoming inputs, decisions are made using only recent (historical) data; we tune based on the past \{20, 40, 80\} batches of inputs and select the one that performs best on the upcoming data. As shown in Figure~\ref{fig:optimal_comparison}, \name{}'s median latency savings are within 4.9--17.2\% of this (more) realistic optimal strategy. %

\para{Comparisons with two-layer inference.} We compare \name{} with in-house implementations of Tabi~\cite{tabi} (NLP) and FilterForward~\cite{canel2019scaling} (CV), which employ compressed models on all inputs, and pass only those inputs with low confidence results to the base model. Acceptable confidence scores from compressed models are configured to ensure both systems operate within the same accuracy loss budget as \name{}. Note that our evaluation is favorable to these systems: we ignore overheads from hosting the compressed models, compute overheads for data pruning (e.g., word pruning in Tabi), and queuing for batch formation between compressed and base models. As shown in Figure~\ref{fig:baseline}, across 3 representative workloads, \name{} delivers 5.7--66.6\% and 20.9--42.0\% lower median and P95 latencies compared to these baselines. For easy inputs, \name{} delivers latency wins by enabling ramps early enough (i.e., within the first third) in the base model to run faster than the baselines' compressed models. For hard inputs, tail latencies with \name{} are capped by its 2\% budget; in contrast, the baselines add the entire compressed model runtime per input.

\begin{figure}[t]
    \centering
    \begin{subfigure}[b]{0.45\linewidth}  
        \centering 
        \includegraphics[width=\linewidth]{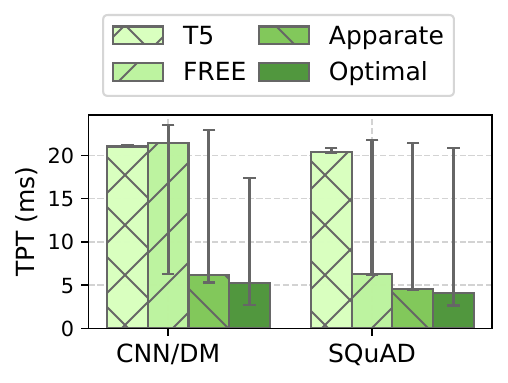}
        \vspace{-22pt}
        \caption{T5}    
    \end{subfigure}
    \hfill
    \begin{subfigure}[b]{0.54\linewidth}
        \centering
        \includegraphics[width=\linewidth]{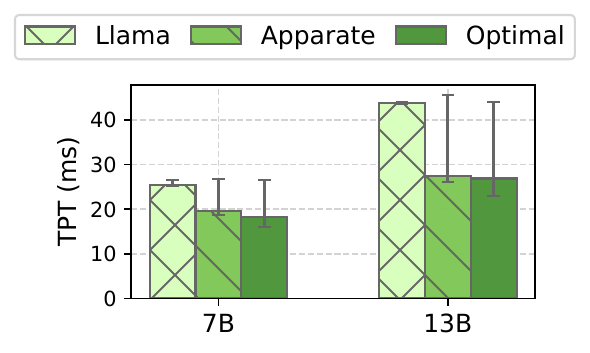}
                \vspace{-22pt}
        \caption{Llama2}
    \end{subfigure}
    \vspace{3pt}
    \tightcaption{Comparison of \name{} against T5 generative LLM and FREE~\cite{FREE} (left), and against Llama generative LLM and its optimal variant (right). Bars represent medians with 25-95th percentiles.}
    \label{fig:llm}
\end{figure}

\para{Varying SLOs.} We considered SLOs for each model that were 2$\times$ and 4$\times$ those in our default experiments (Table~\ref{t:slos}). Generally, higher SLOs induce larger serving batch sizes and higher per-request queuing delays; this dampens \name{}'s relative latency savings (Figure~\ref{fig:slo_sweep}) which target model runtimes, but not queuing delays. For instance, median latency savings for GPT-2 drop from 13.2\% to 6.8\% as SLO grows by 4$\times$. Note that, to illustrate this trend for CV workloads, we upsampled each video to 120 fps. The reason is that our serving platforms are work-conserving and, at 30 fps, they are able to consistently schedule jobs with batch size 1 and low queuing delays given the low model runtimes.

\subsection{Results for Generative Scenarios}
\label{ss:sota_llm}

Figure~\ref{fig:llm} compares \name{} with the vanilla T5-large LLM across our workloads. As shown, \name{} lowers median and 25th percentile TPT by 70.4--77.9\% and 74.6--78.0\%, respectively. However, at the 95th percentile, \name{} incurs 2.6--8.4\% higher TPT due to ramp overheads and (more so) added delay from parallel decoding (non-exiting tokens run alongside the remaining computation for previously-exited tokens). Unlike with classification, formally bounding tail overheads online is hard because variable sequence lengths make TPT values unpredictable in vanilla generative serving. We also evaluated \name{} on the larger Llama models for question answering. Figure~\ref{fig:llm} shows that \name{} reduces the median and 25th percentile TPT by 22.6--37.4\% and 25.4--40.3\%. Notably, the latency wins increase as model size grows.

\para{Comparison to optimal.} Similar to classification, we define the optimal EE strategy as exiting each token at the earliest possible ramp that generates the correct value (ignoring delays for generating the remaining KV states). \name{}'s median TPT for T5 and Llama is within 9.0--16.2\% and 2.1\%--7.7\% of the optimal, respectively. The smaller gap relative to NLP classification stems from two elements. First, accuracy here is measured at the sequence level, granting more flexibility for exiting decisions at individual tokens. Further, auto-regressive generation (with shared state across tokens) grants more continuity than between requests in NLP classification, boosting adaptation benefits.

\subsection{Comparison with Existing EE Strategies}
\label{ss:sota}

\begin{table}[t]
\footnotesize
  \centering
  \begin{tabular}{|c|c|c|c|}
    \hline
    \textbf{} & \textbf{Avg Acc} & \textbf{Median Wins} & \textbf{P95 Wins} \\
    \hline
    \name{} (ResNet50) & 99.0--99.2\% & 46.6--88.6\% & -1.6--0.0\% \\
    BranchyNet & 85.8--99.8\% & -11.0--88.3\% & -11.0\% \\
    BranchyNet+ & 76.1--99.9\% & -11.0--88.3\% & -11.0\% \\
    BranchyNet-opt & 99.0--99.7\% & -11.0--74.5\% & -11.0\% \\
    \hline
    \name{} (BERT-base) & 99.1--99.3\% & 13.7--14.7\% & 2.1--3.0\% \\
    DeeBERT & 91.7--97.1\% & 13.2--36.1\% & -1.3--6.4\% \\
    DeeBERT+ & 82.2--90.3\% & 31.7--36.1\% & 5.9--6.4\% \\
    DeeBERT-opt & 99.0\% & 9.8--36.1\% & -1.4--6.4\% \\
    \hline
  \end{tabular}
  \vspace{8pt}
  \tightcaption{Comparison with existing EE models. Results list accuracies and latency wins for all CV (top) or NLP (bottom) workloads. `+' and `opt' use the optimized tuning from \S\ref{ss:sota}.}%
  \label{t:ee_sota}
\end{table}

\para{Classification.} We compare \name{} with two off-the-shelf EE models: BranchyNet~\cite{branchynet} and DeeBERT~\cite{deebert}. BranchyNet extends ResNet models with ramps of the same style as \name{}, while DeeBERT extends BERT-base with deeper ramps (using the entire BERT pooler - \S\ref{ss:bootstrap}). For each, we follow their prescribed architectures, with always-on ramps after every layer. We perform one-time tuning of thresholds as recommended by both works, and consider two variants: the default recommendation where all ramps use the same threshold, and a version that removes this restriction (\emph{+}). %

Table~\ref{t:ee_sota} presents our results. The main takeaway is that existing EE approaches, even when favorably tuned, yield unacceptable drops in average accuracy up to 23.9\% and 17.8\% for CV and NLP. In contrast, \name{} consistently meets the imposed accuracy constraint (1\% in this experiment) for both workloads. Further, even with such accuracy violations, tail latencies are 0.9--9.4\% lower with \name{} than these systems. The reason is lack of adaptation: all ramps are always active despite current efficacy, yielding undue overheads for non-exiting inputs. In contrast, throughout these experiments, despite having a full ramp budget (for fair comparison), \name{} maintained only 9.1--27.2\% of all possible ramps. %

For fair median latency comparison, we consider an optimally-tuned (opt) version of existing EE models that perform one-time tuning on the actual test dataset, picking the best (latency-wise) thresholds that ensure $<$1\% accuracy drop. As shown, due to its regular and less-constrained adaptation, \name{} outperforms even this oracle version of existing EEs with up to 14.1\% higher median latency savings.

\para{Generative.} 
We compare \name{} with FREE~\cite{FREE}, a state-of-the-art EE solution for T5-large that succeeded CALM~\cite{CALM}. FREE relies on a single, fixed ramp and fine-tunes the entire model to cater to that ramp. The ramp location and weights are selected once to maximize latency savings subject to a 1\% accuracy constraint on a representative dataset (default to the first 3\% of samples). Both FREE and \name{} use T5's final decode head as the ramp architecture. Moreover, to manage tail latencies, \name{} (like FREE) uses a ramp budget of 1. The main difference is that \name{} dynamically adjusts the ramp position (and threshold) to maximize latency savings and preserve accuracy; FREE's omission of runtime adaptation yields 5.5\% accuracy losses in these experiments.

Despite FREE's accuracy violations, \name{} brings median and 25th percentile TPT savings of 28.0--71.0\% and 15.7--28.0\% over FREE (Figure~\ref{fig:llm}). 95th percentile latencies with \name{} are 1.8--2.4\% lower than with FREE because \name{} regularly flushes a batch decoding once the ramp accumulates a pre-specified number of exited tokens. %

\subsection{Microbenchmarks}
\label{ss:deepdive}

Results here use ResNet50 and GPT2-medium running on a random video and Amazon reviews for classification. Reported trends hold for all our workloads.

\begin{figure}[t]
    \centering
    \begin{subfigure}[b]{0.75\linewidth}
        \centering
        \includegraphics[width=\linewidth]{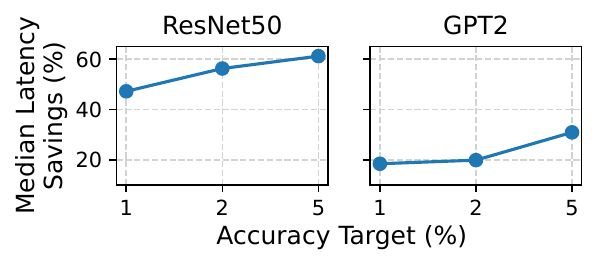}
    \end{subfigure}
    \tightcaption{\name{}'s wins for different accuracy constraints.}  
    \vspace{10pt}
    \label{fig:acc_sweep}
    \label{ss:params}
\end{figure}

\begin{table}[t]
\footnotesize
  \centering
  \begin{tabular}{|c|c|c|}
    \hline
    \textbf{Ramp Budget} & \textbf{ResNet50} & \textbf{GPT2} \\
    \hline
    2\% & 48.9\% & 18.5\% \\
    \hline
    5\% & 49.6\% & 22.2\% \\
    \hline
    10\% & 50.4\% & 24.9\% \\
    \hline
  \end{tabular}
  \vspace{8pt}
  \tightcaption{\name{}'s median latency wins vs. ramp budget.}
  \label{t:ramp_sweep}
\end{table}

\para{Parameter sensitivity.} Recall that \name{} ingests values for two key parameters: ramp aggression/budget and accuracy constraint. Figure~\ref{fig:acc_sweep} and Table~\ref{t:ramp_sweep} studies the effect that these parameters have on \name{}'s latency wins. The findings are intuitive: \name{}'s latency savings over vanilla models decrease as ramp budgets shrink or accuracy constraints tighten. Both trends are a result of \name{} being granted less flexibility for adaptation. Importantly, accuracy constraint has a larger impact on \name{}'s wins. The reason is that inter-ramp dependencies result in overlap in the set of inputs that can exit at any ramp when run in isolation; thus, wins from using more ramps eventually hits diminishing returns. %

\para{Ramp architectures.} Although \name{} opts for using many lightweight ramps, it's adaptation algorithms can support any ramp architecture. To illustrate this, we ran \name{} with DeeBERT's more expensive ramps (\S\ref{ss:sota}). Overall, we find that these costlier ramps dampen \name{}'s latency savings by 4\% since they constrain \name{}'s runtime adaptation in terms of feasible configurations, i.e., fewer active ramps at any time. Crucially, we note that accuracy constraints were still entirely met due to \name{}'s frequent threshold tuning.

\begin{table}[t]
\footnotesize
  \centering
  \begin{tabular}{|c|c|c|}
    \hline
    \textbf{System\textbackslash{}Workload} & \textbf{ResNet50} & \textbf{GPT2} \\
    \hline
    Clockwork & (20.2, 37.8) & (689.2, 779.4) \\
    \hline
    TF-Serve & (24.5, 37.8) & (709.3, 793.1) \\
    \hline
  \end{tabular}
  \vspace{8pt}
  \tightcaption{\name{} on different serving platforms. Results show (median, p95) latency over vanilla models in ms.}
  \label{t:platform}
\end{table}

\para{Impact of serving platform.} \name{} runs atop existing serving platforms, responding to serving and exiting patterns rather than altering platform decisions, e.g., queue management. Table~\ref{t:platform} shows that, despite platform discrepancies, \name{}'s performance wins are largely insensitive to the underlying platform when CV or NLP workloads are configured with the same SLO goal. For example, median latency savings for the Amazon workload and GPT-2 are within 2.9\% when using Clockwork or TF-Serving.

\para{Profiling \name{}.} Figure~\ref{fig:tt_performance} analyzes the run time and optimality of \name{}'s threshold tuning algorithm. Beyond that, \name{} includes two other overheads while running: ramp adjustment and communication between its CPU controller and GPUs for runtime monitoring. Ramp adjustment rounds take an average of 0.5 ms. Coordination overheads are also low (and non-blocking for serving - \S\ref{s:design}) because of \name{}'s small ramp sizes (definitions and weights consume $\sim$10KB) and profiling data (simply a top-predicted result with an error score, collectively $\sim$1KB). In total, CPU-GPU coordination delays take an average of 0.5ms per communication, 0.4ms of which comes from fixed PCIe latencies in our setup.

\para{Importance of \name{}'s techniques.} \name{}'s runtime adaptation considers frequent (accuracy-guided) threshold tuning, with periodic ramp adjustments. Table~\ref{t:ee_sota} shows the importance of threshold tuning on average accuracies. We also evaluate the importance of ramp adjustment on \name{}'s latency wins by comparing versions with and without it. Overall, disabling ramp adjustment results in 20.8--33.4\% lower median latency wins, though worst-case latency (and throughput) and accuracy constraints remain continually met.

%% file: related.tex
\section{Additional Related Work}
\label{s:related}

\para{Early exit networks.} Existing proposals focus on exit \emph{ramp architecture} and \emph{exit strategy}~\cite{deebert,branchynet,msdnet,berxit,fastbert,RightTool,pabee,FREE,CALM} for specific models. Ramp architectures are domain-specific, but replicating the last (few) layers is the common practice~\cite{deebert,berxit}; \name{} builds on this and prefers shallow ramps (\S\ref{s:design}). Existing exit strategies consider \emph{confidence} of the labels~\cite{fastbert}, \emph{entropy} of the prediction~\cite{deebert}, or more sophisticated elements such as counters across ramps~\cite{pabee}. \name{} is agnostic to EE technique, and instead focuses on bringing EEs to arbitrary models and providing runtime management to maximize latency savings and accuracy preservation.

\para{Model optimizations.} Much work has focused on creating \emph{variants} of ML models to optimize serving. Some employ compiler techniques to analyze (and optimize) execution at a graph or operator level for lower latency~\cite{tensorrt,tvm}. Other techniques compress models into versions with fewer layers or less weight specificity while adhering to accuracy objectives, e.g., quantization~\cite{gptq,squeezellm,awq}, distillation~\cite{incontextdistill,distilbert}, and model pruning~\cite{sparsegpt,llmpruner}. These efforts are complementary to \name{}, which would bring input-level compute adaptation to their outputs, i.e., on optimized or compressed models. Moreover, unlike many of these techniques, \name{} does not alter the original model's weights and instead preserves its full predictive power for hard inputs and constant feedback.

\para{Inference optimizations.} Recent work optimizes model serving objectives based on workload characteristics~\cite{inferline,infaas,shepherd,swayam,nexus,lazybatching,mark}. For instance, Inferline~\cite{inferline} optimizes serving cost while adhering to strict latency constraints using intelligent provisioning and management. Shepherd~\cite{shepherd} maximizes goodput and resource utilization by leveraging cross-workload predictability. Despite their impressive results, these works optimize their metric of choice at the expense of latency and do not resolve the latency-throughput tension, which is the focus of our (complementary) work. Mystify~\cite{mystify} and INFaaS~\cite{infaas} generate and choose model variants based on their intent and constraints (including performance). As noted above, \name operates on the output of such tools, bringing latency wins to compressed models (\S\ref{ss:main_results}).

\para{Dynamic neural networks.} Other techniques boost inference efficiency by adapting model execution~\cite{skipnet,switchtransformer,multiscale,dynamicseg, modulating} at different granularities, e.g., sample-wise, spatial-wise, and temporal-wise. As with exiting, their modulations to serving risk accuracy violations and foregone latency wins with dynamic workloads. We hope that \name{} can motivate and guide adaptation systems to navigate accuracy-latency-throughput tradeoffs for such techniques, but leave that to future work. Other low-level (e.g., GPU kernel level) techniques that optimize execution of dynamic neural networks~\cite{Brainstorm,cocktailer} can benefit \name{} and improve its performance.

%% file: conclusion.tex
\section{Conclusion}
\label{s:conclusion}

We present \name{}, the first system that automatically injects and manages early exiting for ML inference. Key to \name{}'s ability to alleviate latency-throughput tensions in serving is its use of exiting only for fast results (not compute savings). This provides continual feedback on exits, and powers \name{}'s novel adaptation strategies for EE ramps and thresholds. \name{} lowers median latencies by 40.5--91.5\% and 10.0--24.2\% for diverse CV and NLP workloads, and reduces median time-per-token latencies by 22.6--77.9\% for generative workloads, all while meeting accuracy constraints and preserving platform throughputs.

\para{Acknowledgements.} We thank Wyatt Lloyd, the SOSP reviewers, and our shepherd, Rong Chen, for their helpful and constructive comments. This work was supported by a Sloan Research Fellowship and NSF CNS grants 2147909, 2151630, 2140552, 2153449, and 2152313.

%% file: appendix.tex
\section{Appendix}
\label{s:appendix}

The appendix was not peer-reviewed.

\noindent We list the SLOs for each model in the classification experiments (Figures~\ref{fig:cv_med}--\ref{fig:nlp_cdf}). The default SLOs for classification models are $2\times$ each model's inference time with batch size 1.

\begin{table}[h!]
\small
  \centering
  \begin{tabular}{|c|c|c|}
    \hline
    \textbf{Model} & \textbf{Latency w/ bs=1 (ms)} & \textbf{Default SLO (ms)} \\
    \hline
    ResNet18 & 6.5 & 13.0\\
    \hline
    ResNet50 & 16.4 & 32.8\\
    \hline
    ResNet101 & 33.3 & 66.6\\
    \hline
    VGG11 & 3.3 & 10.0\\
    \hline
    VGG13 & 3.8 & 10.0\\
    \hline
    VGG16 & 4.5 & 10.0\\
    \hline
    Distilbert-base & 15.5 & 31.0\\
    \hline
    BERT-Base & 29.4 & 58.8\\
    \hline
    BERT-Large & 63.2 & 126.4\\
    \hline
    GPT2-medium & 103.0 & 206.0\\
    \hline
  \end{tabular}
  \caption{Different SLOs used in \S\ref{ss:main_results} and Figure~\ref{fig:slo_sweep}. All numbers are measured on the A6000.}
  \label{t:slos}
\end{table}

\noindent We also list the pseudocode for our threshold tuning and ramp adjustment algorithms described in the design section. %

\begin{algorithm}[b]
\small
\DontPrintSemicolon
  \KwInput{$ramps$, list of ordered active ramp IDs}
  \KwInput{$acc\_loss\_budget$, max accuracy loss tolerable}
  \KwInput{$smallest\_step\_size$, smallest step size for incrementing thresholds}
  \KwOutput{$thresholds$, thresholds associated with each ramp}
  \KwOutput{$latency\_savings$, latency savings with searched thresholds}
  \tcc{all thresholds start at 0, i.e. no EE}
  $thresholds\gets [0.0] * len(ramps)$

  \tcc{each ramp has its own step size}
  $step\_sizes\gets [smallest\_step\_size] * len(ramps)$
  \label{alg:initialization}

  \While{True}
    {
        $best\_ramp, overstepped\_ramps, latency\_savings \gets pick\_ramp(ramps, thresholds, acc\_loss\_budget)$
        \tcc{find next ramp to update thresholds} \label{alg:pick_ramp}
        \If{$best\_ramp$ is valid}
        {
            \tcc{increment threshold of the selected ramp} 
            $thresholds[best\_ramp] \mathrel{+=} step\_sizes[best\_ramp]$ \\
            \tcc{double step\_size}
            $step\_sizes[best\_ramp] \mathrel{*}= 2$ \label{alg:double_step}
        }
        \Else
        {
            \If{$step\_sizes.all() \leq smallest\_step\_size$} 
            {
                \Return $thresholds$, $latency\_savings$  \label{alg:end}
            }
        }
        \tcc{half step\_size for overstepped ramps, double step\_size for the rest}
        \For{$ramp$ in overstepped ramps}
        {
            $step\_sizes[ramp] \mathrel{/=} 2$ \label{alg:shrink_step}
        }
    }
\caption{Threshold tuning algorithm}
\label{alg:threshold_search}
\end{algorithm}

\begin{algorithm}[b]
\small
\DontPrintSemicolon
  \KwInput{$P$, latest positive ramp}
  \KwInput{$ramps$, list of ramp ids}

  \tcc{Initialize ramp settings}
  \tcc{Check whether negative ramps exist}
  \If{negative utility ramps exist}{
    Try fast threshold tuning\;
    \If{still negative}{
      Deactivate all negative-utility ramps\;
    } \Else{
      Update thresholds\;
      \Return
    }
     \tcc{Determine candidate ramps located at the middle of each interval after $P$ (see Figure~\ref{fig:two_ramps})}
      $candidates \gets pick\_candidates(P, ramp\_ids)$ 
    
      \tcc{Compute utilities and select ramp with best upper-bound utility (see Figure~\ref{fig:two_ramps})}
      \ForEach{$candidate$ in $candidates$}{
        Compute upper-bound exit rate\;
        Calculate utility using upper-bound exit rate\;
        \If{utility is positive and utility $>$ $selectedRamp$.utility}{
          $selectedRamp \gets candidate$\;
        }
      }
      \tcc{Update EE configuration}
      Remove all deactivated ramps\;
      Add $selectedRamp$ with initial threshold=0\;
      Update thresholds in future tuning rounds\;
  }

  \tcc{If all ramps have positive utilities, explore the possibility of higher latency savings}
  \If{all utilities positive}{
    \If{ramp budget allows}{
      Add a ramp before the highest utility ramp\;
    }
    \Else{
      Shift the lowest utility ramp one position earlier\;
    }
  }

  \Return{}\;
\caption{Ramp adjustment algorithm}
\label{alg:ramp_adjustmemt}
\end{algorithm}